\documentclass[aps,pra,reprint,twocolumn,showpacs,floatfix,superscriptaddress]{revtex4-2}

\usepackage{amssymb,amsmath,amstext}                
\usepackage{graphicx}                                               
\usepackage{epstopdf}                                               
\usepackage{color}                                                     
\usepackage{bm}                                                        
\usepackage{appendix}                                              
\usepackage[utf8]{inputenc}
\usepackage{bbold}
\usepackage{bbm}
\usepackage{ulem}[normalem]
\usepackage{latexsym}
\usepackage{xcolor}
\usepackage{braket}
\usepackage{mathtools}
\definecolor{lblue} {RGB}{51,71,158}
\usepackage[colorlinks=true,citecolor=blue,linkcolor=blue,urlcolor=lblue]{hyperref}

\DeclareMathOperator{\Tr}{Tr}

\newcommand{\be}{\begin{equation}}
\newcommand{\ee}{\end{equation}}

\begin{document}

\title{Many-body localization in tilted and harmonic potentials}

\author{Ruixiao Yao}
\affiliation{Department of Physics and State Key Laboratory of Low Dimensional Quantum Physics, Tsinghua University, Beijing, 100084, China}
\author{Titas Chanda}
\affiliation{Institute of Theoretical Physics, Jagiellonian University in Krak\'ow,  \L{}ojasiewicza 11, 30-348 Krak\'ow, Poland }
\author{Jakub Zakrzewski}
\email{jakub.zakrzewski@uj.edu.pl}
\affiliation{Institute of Theoretical Physics, Jagiellonian University in Krak\'ow,  \L{}ojasiewicza 11, 30-348 Krak\'ow, Poland }
\affiliation{Mark Kac Complex
Systems Research Center, Jagiellonian University in Krakow, Krak\'ow,
Poland. }

\date{\today}

                              
\begin{abstract}
We discuss nonergodic dynamics of interacting spinless fermions in a tilted optical lattice as modeled by XXZ spin chain in magnetic (or electric) field changing linearly across the chain. The time dynamics is studied using exact propagation for small chains and matrix product states techniques for larger system sizes. We consider both the initial N\'eel separable state as well as the quantum quench scenario in which the initial state may be significantly entangled. We show that the entanglement dynamics is significantly different in both cases. In the former a rapid initial growth is followed by a saturation for sufficiently large tilt, $F$. In the latter case the dynamics seems to be dominated by pair tunneling and the effective tunneling rate scales as $1/F^2$. In the presence of an additional harmonic potential 
the imbalance is found to be entirely determined by a local effective tilt, $F_{\text{loc}}$, the entanglement entropy growth is modulated with frequency that   follows $1/F_{loc}^2$ scaling first but
at long time the dynamics is determined rather by the curvature of the harmonic potential.  Only for large tilts or sufficiently large curvatures, 
corresponding  to the deeply  localized regime, 
we find the logarithmic entanglement growth for N\'eel initial state. The same curvature determines long-time imbalance for large $F$ which reveals  strong revival phenomena associated with the manifold of equally spaced states,  degenerate in the absence of the harmonic potential.
\end{abstract}

\maketitle

\section{Introduction}

Many-body localization (MBL) became a well established property of interacting many-body systems in sufficiently strong disordered potentials.  Seminal works \cite{Gornyi05,Basko06} started a chain of research resulting in hundred of research papers (to see some of the early works, consult \cite{Znidaric08,Santos10}) partially summarized in recent reviews  \cite{Huse14, Nandkishore15, Alet18, Abanin19}.
MBL is considered as a robust counterexample to thermalization, as localization is supposed to lead to the memory effect of the initial state in the closed system in contrary to the eigenstate thermalization hypothesis \cite{Deutsch91,Srednicki94} predictions.  Yet, a recent work \cite{Suntajs20e} put the very existence of MBL in the thermodynamic limit in question stimulating an intensive debate  \cite{Panda20,Sierant20,Suntajs20,Laflorencie20,Sierant20p,Abanin21} mainly based on results coming from exact diagonalization studies,
necessarily limited to small system sizes. 
Matrix product states (MPS)  techniques allow one to consider large system sizes \cite{Vidal03,White04, Daley04,Haegeman11,Schollwoeck11,Paeckel19}. Then the transition to MBL is studied addressing long time dependence of certain correlation function (the so called imbalance, see below) \cite{Sierant17,Sierant18,Zakrzewski18,Doggen19,Doggen20,Chanda20} 
mimicking experimental scenarios \cite{Schreiber15,Luschen17}.  Still these claims are not definitely conclusive due to finite systems sizes and, importantly,  rather short evolution times taken into consideration, limited by the numerical methods, {such as the}   time-dependent variational principle (TDVP).

{Recently, it was also observed that a}
 constant strong tilt of the optical lattice may lead to nonergodic behavior in the presence of a small disorder \cite{vanNieuwenburg19} or
 even in the absence of it \cite{Schulz19}. The phenomenon, usually referred to as the  Stark many-body localization (SMBL), is  an extension of Wannier-Stark localization for non-interacting particles \cite{Gluck02}. Interestingly, for a pure constant tilt the dynamics is a bit peculiar, e.g., the
 entanglement entropy shows a strong initial growth for an initial separable state  \cite{Schulz19}, a phenomenon associated to the fact that such an initial state is degenerate with the number of other separable states of the same global dipole moment (see below). On the other hand, with addition of a small disorder, or an additional weak harmonic potential SMBL shows features very similar to a standard MBL such as
 the logarithmic entanglement entropy growth \cite{Schulz19}, Poissonian level spacing statistics \cite{vanNieuwenburg19,Schulz19},  a 
 power-law decay of a double electron-electron resonance (DEER) (introduced by \cite{Serbyn14}) signal  or a logarithmic growth of quantum mutual information \cite{Taylor20}. 
 The role of the harmonic confinement was further analyzed in \cite{Chanda20c} where it was shown, in particular, that a smooth site-dependent chemical potential leads to an effective local static field related to the derivative of this potential with respect to the spatial dimension.  For strong harmonic term that may lead to phase separation between localized and delocalized phases \cite{Chanda20c,Yao21}. To complete the picture, let us also note  that SMBL is also addressed recently for open systems \cite{Wu19}.

  While spinless fermions have been considered theoretically (bosons have been discussed in \cite{Taylor20,Yao20b}), the experiments
 with cold atoms considered spinful fermions both in two-dimensional (2D) setup \cite{Guardado20} and in a single dimension (1D) \cite{Scherg20}. SMBL has been also  studied on superconducting quantum processor \cite{Guo20} emulating a spin-1/2 chain in a triangular ladder as well as, very recently,  on an {trapped} ion quantum simulator \cite{Morong21}, where the coexistence of localized and delocalized phases in a strong harmonic potential \cite{Chanda20c} has been verified. 
 
 The unusual time dynamics of pure tilted case (by a pure case we shall consider the situation with the ideal linear lattice tilt with no additional harmonic or other local potential) may be better understood using the concept of the emergent dipole moment conservation \cite{Guardado20}. Due to superextensive character of the tilted potential term, the energy accumulated in the potential energy related to the tilt may not, sometimes change entirely into the kinetic energy leading to the memory of the initial state. This point has been further elaborated in \cite{Khemani20} using the concept of shattered Hilbert space. State possessing initially a large (in absolute value) dipole moment may not undergo thermalization. This has been visualized in \cite{Doggen20s} for domain wall states also for a relatively small tilt when the majority of separable initial states thermalizes fast. Their existence may be also  traced back to a peculiar feature of the Heisenberg model used -- the domain containing spins with the same orientation (up or down oriented spins) is practically frozen in the dynamics for the time dependent on the length of the domain.  So local thermalization occurs at domain walls only \cite{Doggen20s, Yao21}.
 
 The aim of this contribution is to provide a detailed insight into the dynamics of both pure tilt case as well as a harmonically perturbed model. 
 We consider small system sizes with large integration times together with larger systems treated using time dependent variational principle (TDVP) algorithm \cite{Haegeman11,Haegeman16,Paeckel19}. We analyze in detail different contributions to the entanglement entropy and their time dynamics. Importantly, apart from the dynamics of initial separable N\'eel state, we consider  also  the quantum quench scenario \cite{Naldesi16,Kubala21} for a pure tilt case used previously for disordered systems. That allows us to provide insight into the long time entanglement growth. In the presence of an additional harmonic trap we verify the transition to a more standard, MBL-like behavior \cite{Schulz19,Taylor20} although we observe a logarithmic growth of the entanglement entropy in the deeply localized regime only.
{Most importantly, we stress the importance of the concept of \textit{local} tilt, i.e., the spatial derivative of the continuously varying chemical potential.
Specifically, we show that the dynamics can be described entirely by the local tilt in cases of both pure tilt and with the addition of harmonic potential when viewed through the eye of time-evolved \textit{local} imbalance. This is due to the fact that a particular site is only affected by few neighboring sites during the evolution because of strong localization rendering the dynamics very much local in space.
On the other hand, as mentioned before, addition of harmonic potential greatly changes the dynamics of entanglement
entropy. However, we show that the concept of local tilt still remains valid for entanglement
entropy within the scenarios of harmonically perturbed systems.}
{Furthermore}, for sufficiently large tilt, we provide an evidence of quantum revivals present in the (easily accessible to experiments) imbalance as well as reflected in the entropy dynamics.

\section{The model}

We consider the paradigmatic model in MBL studies, namely the XXZ spin chain with the Hamiltonian 
\begin{equation}
 H= J\sum_{l=1}^{L-1} \ (S^x_lS^x_{l+1}+S^y_lS^y_{l+1}) +V\sum_{l=1}^{L-1}S^z_lS^z_{l+1} + \sum_{l=1}^{L}  \mu_lS^z_l,
 \label{eq: XXZ}
\end{equation}  
with $ S^i_l $'s being  spin-1/2 operators and $\mu_l$  being the onsite potential.  In typical MBL studies \cite{Luitz15,Alet18,Sierant19b,Sierant20},  $\mu_l$ is a diagonal disorder (a magnetic field along $z$-axis) drawn from random uniform distribution.  We consider instead $\mu_l=Fl+Al^2/L^2$ where $F$ is the magnitude of the linear tilt and  $A$ is (twice) the curvature of the additional harmonic potential.  Note that while for $F=0$ case it may be convenient to center the harmonic potential with respect to the chain (so it forms a confining potential \cite{Chanda20c, Yao20b}). Here we follow a different setting \cite{Schulz19,Taylor20}.  We also  set $J=1$ as the energy unit.
For $V=J$ the model reduces to the Heisenberg chain. Using Jordan-Wigner transformation the model reduces to the case of spinless fermions tunneling with the rate  {$J/2$} and with the nearest neighbor interaction $V$. {Unless explicitly stated otherwise, we shall mostly concentrate on the dynamics of the Heisenberg chain.}

We mostly consider time dynamics starting from the initial N\'eel state with every odd spin pointing up,  and every even down (remaining in the conserved total $S^z=0$ sector). While Ref. \cite{Schulz19} averaged the dynamics over different possible states, we choose the N\'eel state 
that corresponds,  after Jordan-Wigner transformation, 
to an equivalent density wave state with occupied odd sites and empty even sites in the spinless fermion representation in analogy to the experimental situation \cite{Luschen17}.

We characterize the state of the system by considering the spin correlation function which,  for the N\'eel state,  is equivalent to the imbalance of occupations between odd and even sites in the fermionic picture:
\be
I(t)=\frac{N_o(t)-N_e(t)}{N_o(t)+N_e(t)}=\frac{4}{L}\sum_{l=1}^L \langle S^z_l(t)\rangle\langle S^z_l(0)\rangle,
\label{eq:imb}
\ee
where $N_i,\ i=e,o$, denote the total occupation of even and odd sites.  We use also the entanglement entropy by splitting the system in two parts $A$ and $B$.  Defining the reduced density matrix $\rho_A=\Tr_B |\psi\rangle\langle \psi|$ where $|\psi(t)\rangle$ is the time evolved state of the system, the entanglement entropy reads $S=-\sum_i \lambda_i \ln(\lambda_i)$ where $\lambda_i$ are the eigenvalues of  $\rho_A$.  Interestingly, 
{owing to the global $U(1)$ symmetry of total $\sum_l S^z_l$ conservation,}
the entanglement entropy may be split into the sum of classical number entropy $S_N$ and the quantum configuration entropy $S_C$: $S=S_N+S_C$ \cite{Schuch04,Schuch04b,Donnelly12,Lukin19,Sierant19c} with
\begin{eqnarray} 
S_N(t)&=&- \sum_{n=0}^N p_n \log(p_n)\hfil \nonumber \\
S_C(t)&=&- \sum_{n=0}^N  p_n \sum_i \rho_{ii}^{(n)}  \log(\rho_{ii}^{(n)}),
\label{eq:ent}
\end{eqnarray} 
where we denoted  by $p_n$ the probability of populating the $n$-particle sector
in subsystem A and by $\rho^{(n)}$
the corresponding block of the density matrix $\rho$ for the subsystem A,
such that $\rho_A=\sum_n
p_n \rho^{(n)}$. The number entropy  describes a real transfer of particles between the subsystems considered and the configuration entropy, $S_C$, quantifies  the possible arrangements of particles on each side of the system.

\section{Pure tilt case}

Let us consider first the dynamics in the pure tilted case with $A=0$. We shall consider both small systems amenable to exact diagonalization as in \cite{Schulz19,vanNieuwenburg19,Taylor20} as well
as more realistic systems sizes using TDVP. To check convergence we have also compared TDVP and TEBD results as in \cite{Chanda20} -- see Appendix \ref{app:mps}. For a purely deterministic potential no averaging over disorder realizations, as in case of random potentials, is possible. 
Typically we shall consider small systems of size $L=16$ and compare those with $L=40$ cases analyzed with TDVP.

\subsection{Initial N\'eel state}

As announced in the previous section we consider the time dynamics of the initial N\'eel state.
 \begin{figure}
 \includegraphics[width=.9\linewidth]{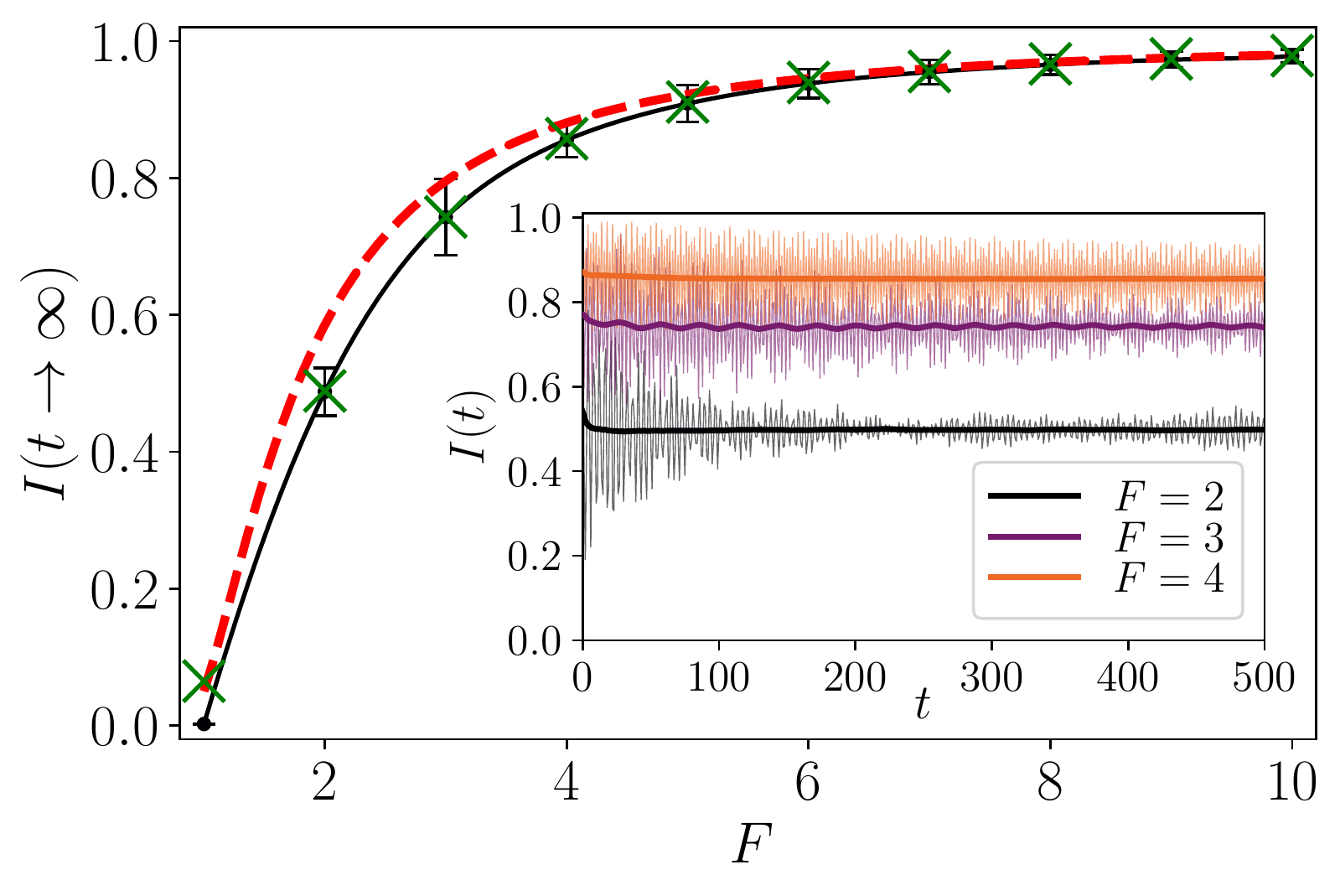}
   \caption{The  late time imbalance (averaged over times in [1500, 2000] interval) as a function of $F$. The imbalance decays to zero around $F=1$. The system size is $L=40$ (black symbols with error bars), while green crosses correspond to $L=16$. The red dashed line gives the analytic prediction for noninteracting case \eqref{wannier}.
The inset presents exemplary  time-dependencies of imbalance for large system-size  for few $F$ values indicated.  It shows strong fast oscillations (related to Bloch oscillations) -- compare Fig.~\ref{fig:imbs} for small system sizes.  The solid lines are obtained  by filtering out large frequencies in the FFT of $I(t)$ signal.  }
 \label{fig:imbalnew1}
\end{figure}
Consider first the ``experimental'' evidence for SMBL, i.e., the existence of a non-zero imbalance at large times indicating the preservation of the memory about the initial state during time evolution. The imbalance at saturation occurring at large times is presented in Fig.~\ref{fig:imbalnew1}. The data are averaged over $t\in[1500,2000]$ to wash out the rapid oscillations of the time-dependent signal (as shown in the inset).  Data are obtained with TDVP with $\chi_{max}=512$. The errors indicated in the figure are due to these fast oscillations and are standard statistical errors. The data indicate that the crossover to delocalized phase appears at $F \approx 1$,  in agreement with the data for small system sizes obtained from level statistics \cite{Schulz19,vanNieuwenburg19}. One should, however, keep in mind that the accuracy of long time TDVP simulations close to the transition deteriorates, so that the critical $F$ value should be taken with caution as the result becomes a compromise between the duration of the time propagation and the limitation on the assumed auxiliary  space dimension,  $\chi_{max}$. Note, however, a very good agreement for $F\ge2$ between long time imbalance value for $L=40$ obtained via TDVP and $L=16$ obtained via exact diagonalization-based propagation. In both cases 4 sites at each edge of the system were excluded from calculation of the imbalance to minimize open boundary effects. Also note, to some extent disappointingly, that the final imbalance is quite close to the analytic prediction for the noninteracting case, i.e., pure Wannier-Stark localization \cite{Scherg20}
\begin{equation}
I_{WS}(F)={\cal J}_0^2\left(\frac{2J}{F}\right),
\label{wannier}
\end{equation}
{where ${\cal J}_0(x)$ represents the zeroth-order Bessel function,} especially for large $F$ that suggests the decreasing importance of  interactions for large $F$.

\begin{figure}[ht]
 \includegraphics[width=0.5\linewidth]{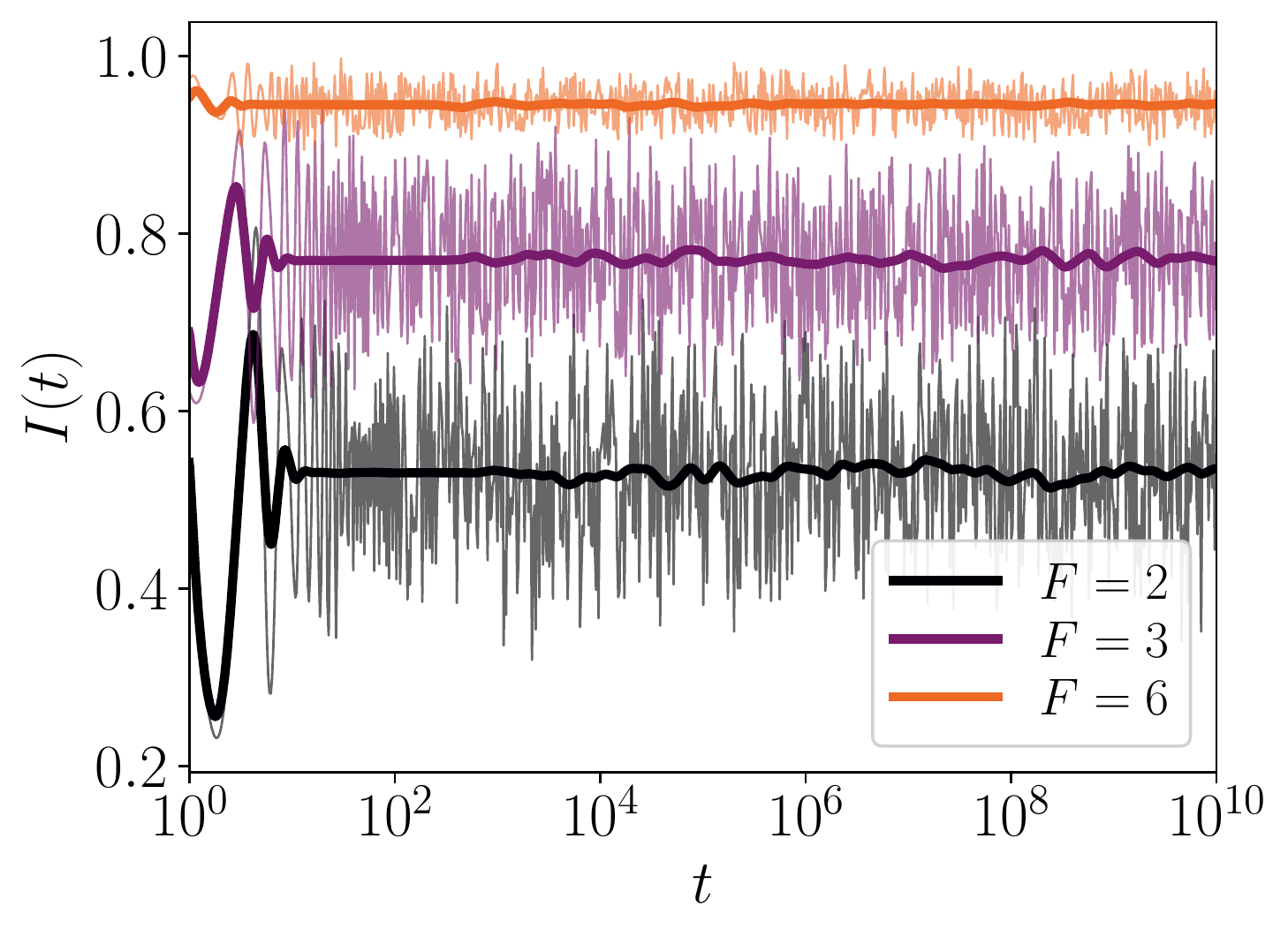}\includegraphics[width=0.5\linewidth]{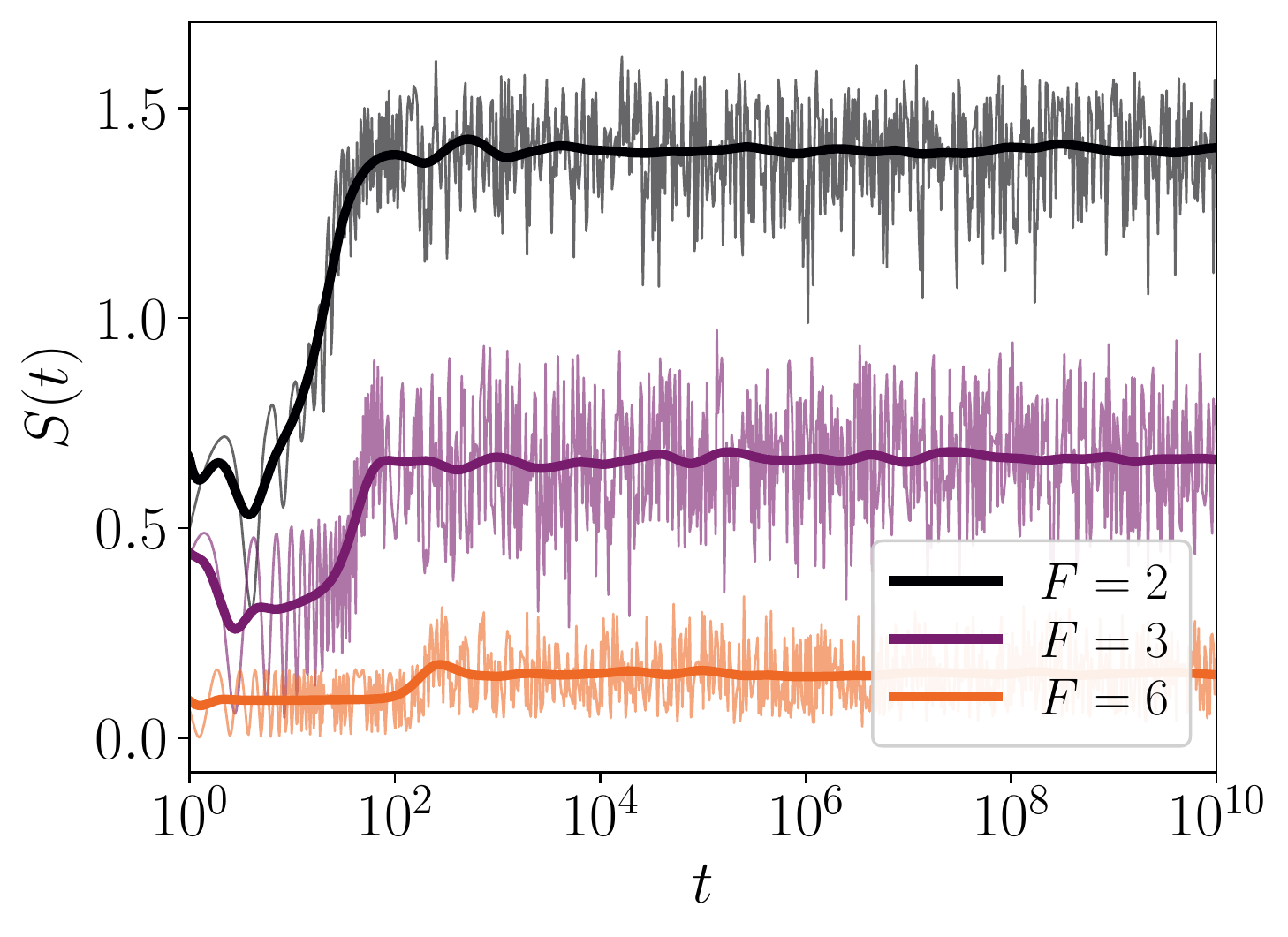}
   \caption{Left: The imbalance as a function of time for $L=16$ for the Heisenberg chain ($V=1$) at few values of the static tilt $F$ as indicated.  The imbalance shows large fast  oscillations -- the thick lines are obtained by filtering out high frequencies in the time signal -- see the text for details. Right: The entanglement entropy between left and right half of the system. Fast initial growth is followed by a saturation.  Thick lines represent the frequency filtered data.}
     \label{fig:imbs}
\end{figure}

 Left panel of Fig.~\ref{fig:imbs} shows a typical imbalance as a function of time for a few values of the static tilt, $F$ for small system size $L=16$ but for much longer times. Observe quite large fast changes of $I(t)$ which reflect short range jumps of fermions (spin flips in the spin language) between nearby sites. Those are the remnants of Bloch oscillations for noninteracting system \cite{Scherg20} in our interacting case. To get the information beyond these rapid changes we apply a frequency filter that effectively averages our fast oscillations. The filter smoothly cuts out large frequencies in the Fourier transform of the signal --
for a detailed description of the filter see the Appendix \ref{app:filter}. The effective, time-averaged signal is also shown in  Fig.~\ref{fig:imbalnew1}. 

The right panel  of Fig.~\ref{fig:imbs} shows the entanglement entropy growth with time. Here an initial rather rapid growth is also followed by a saturation for sufficiently large tilts as taken for the plots. The values of the tilt $F$ correspond to the ``localized'' regime according to spectral properties -- the localization transition is estimated at $F_c\approx 1.1$ {in the large system-size limit}
 \cite{vanNieuwenburg19} (note that different units are used in that work, the value given there is twice bigger).  Let us recall that one of the hallmarks of MBL is the logarithmic growth of the entanglement entropy for initially separable states \cite{Bardarson12,Serbyn13a}.  The growth of the entanglement entropy for SMBL is apparently quite different.
   Moreover, the fast saturation of the entanglement entropy with time is in apparent contradiction with the result shown in \cite{Schulz19} where the entanglement entropy, after a fast initial rise, grows slowly for long times too. This is due, in our opinion, to a different initial state assumed. We take an ideal N\'eel state while in \cite{Schulz19} an average over initial states with the same density per site but having no initial occupation on sites between which the cut into A and B subsystems is made. Thus among states included in \cite{Schulz19} are, e.g., states with domain walls -- extended  regions of identically oriented spins that  may result in excessively slow dynamics 
 due to frozen spins \cite{Doggen20s,Yao21}.
 
 \begin{figure}
 \includegraphics[width=\linewidth]{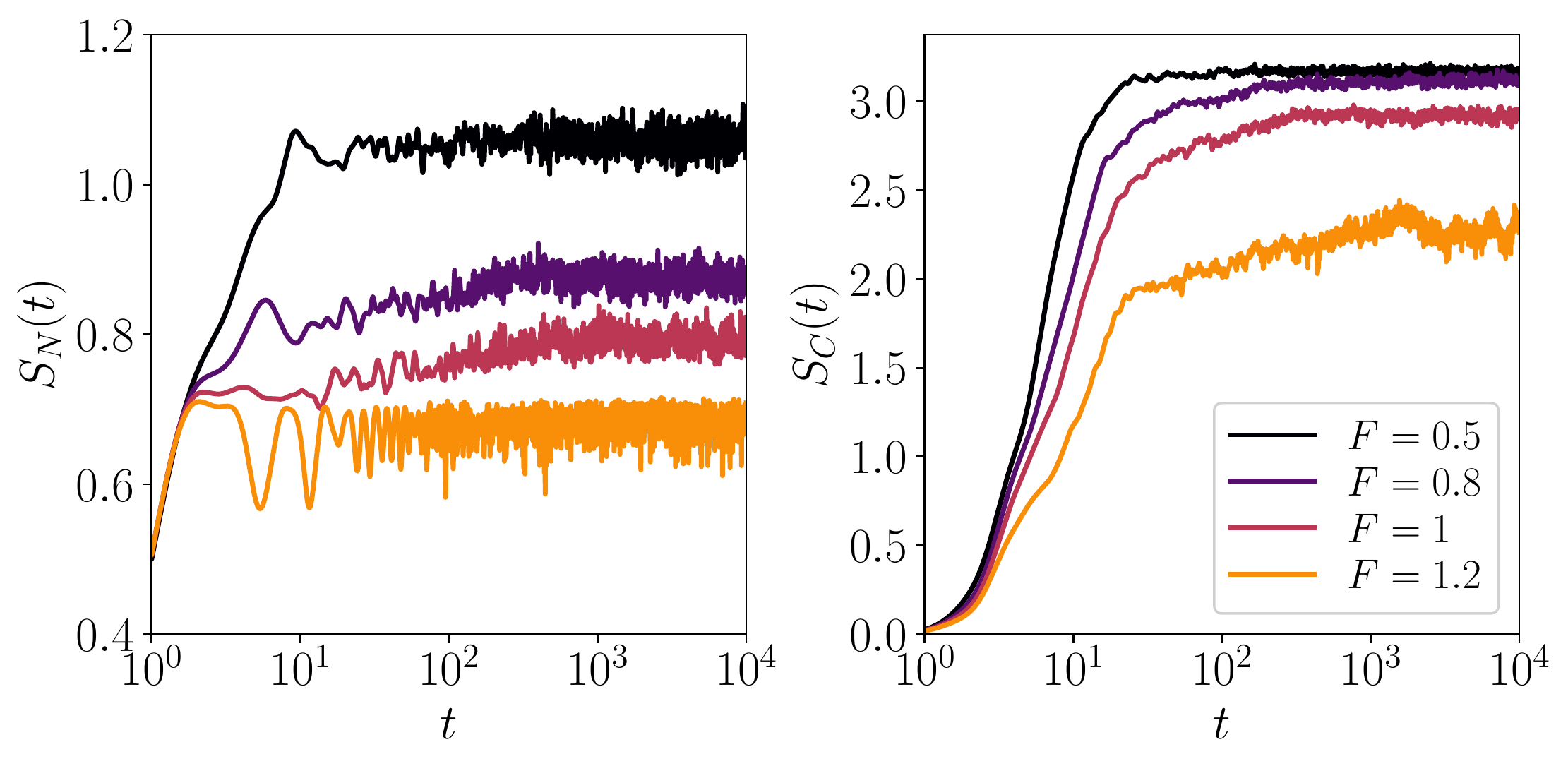}
 \caption{The number (left) and configuration (right) entropies, {defined in   \eqref{eq:ent}},  between left and right half of the system for $L=16$ in time for the initial N\'eel state. For moderate values of the tilt $F\in[0.5,1.2]$  the fast initial growth is followed by a logarithmic growth and, eventually, saturation
 }
 \label{fig:moder}
\end{figure}
While Fig.~\ref{fig:imbs} presents the results for a relatively large tilt, deep in the localized regime,  Fig.~\ref{fig:moder} presents the time evolution of the entanglement entropy for tilt values close to the transition. Here we split the entropy into the classical number entropy (left panel) and the configuration entropy (right panel) as described by \eqref{eq:ent}. Observe that for both entropy components the fast growth and the saturation part are separated by an intermediate stage of logarithmic entropy growth. As pointed out in \cite{Schulz19,Taylor20} the fast initial entropy growth is related to degenerate manifold of states 
  leading to the entanglement build-up. For moderate tilt the localization takes some time to settle in leading to an intermediate logarithmic growth resembling MBL. This region of $F$ values, say $F\in[0.5,1.2]$,  corresponds to the transition region between delocalized and SMBL phases in the spectral analysis \cite{vanNieuwenburg19}.
 
Let us note also that the saturation of the number entropy at low values at later times makes the localization observed distinctly different from disorder induced MBL where a very slow $S_N(t)\sim \ln(\ln t)$ is observed \cite{Kiefer20,Kiefer21}.  Here for MBL also an eventual  saturation is predicted \cite{Luitz20} due to single particle exchange across the analyzed bond but the corresponding saturation value is much higher than observed by us in SMBL (compare Fig.~\ref{fig:moder}).

 \begin{figure}
 \includegraphics[width=.9\linewidth]{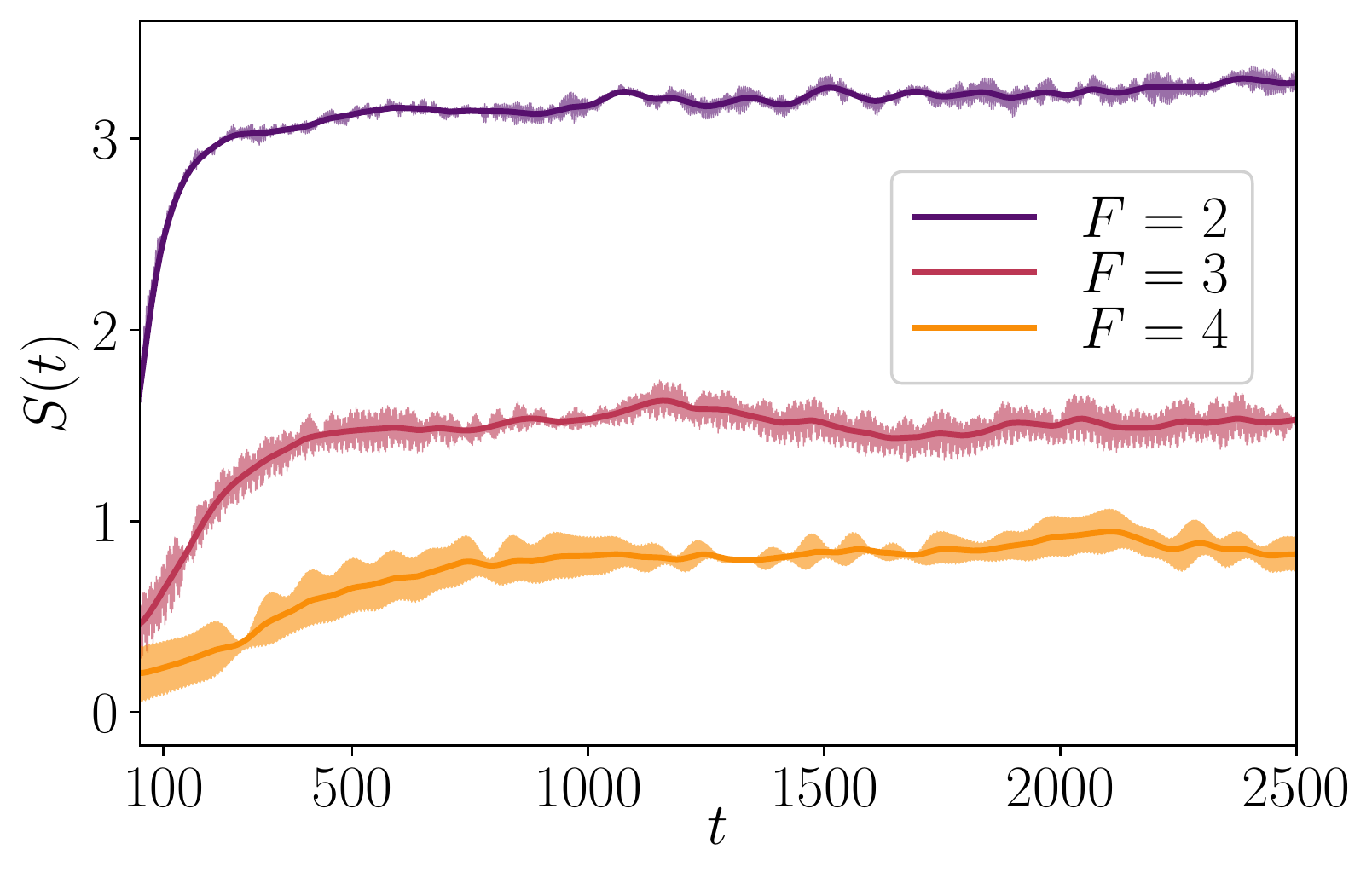}
   \caption{The entanglement entropy time dependence on the bond located in the middle of the system for $L=40$ and few values of $F$ as indicated in the figure. The fast linear initial growth changes into saturation with quite strong fluctuations.
 }
 \label{fig:entrol}
\end{figure}
These observations are further exemplified for larger system sizes using TDVP simulations.
The entanglement entropy growth on the middle bond is shown in Fig~\ref{fig:entrol}. Strong high frequency oscillations can be averaged out leading to solid lines. Two time-regimes may be again clearly identified. The first is the region of a linear growth of entropy with time which extends to larger times for bigger $F$. The second regime,  for sufficiently long times and larger $F$ corresponds to an apparent saturation of the entanglement entropy.  Data for $F=3$ and larger clearly saturate while for lower $F=2$ we observe a fast linear growth for short times followed by a slow linear drift without saturation.  We cannot determine whether this drift is linear or logarithmic in time due to a limited time span and small changes of the entropy.

\begin{figure}
 \includegraphics[width=0.9\linewidth]{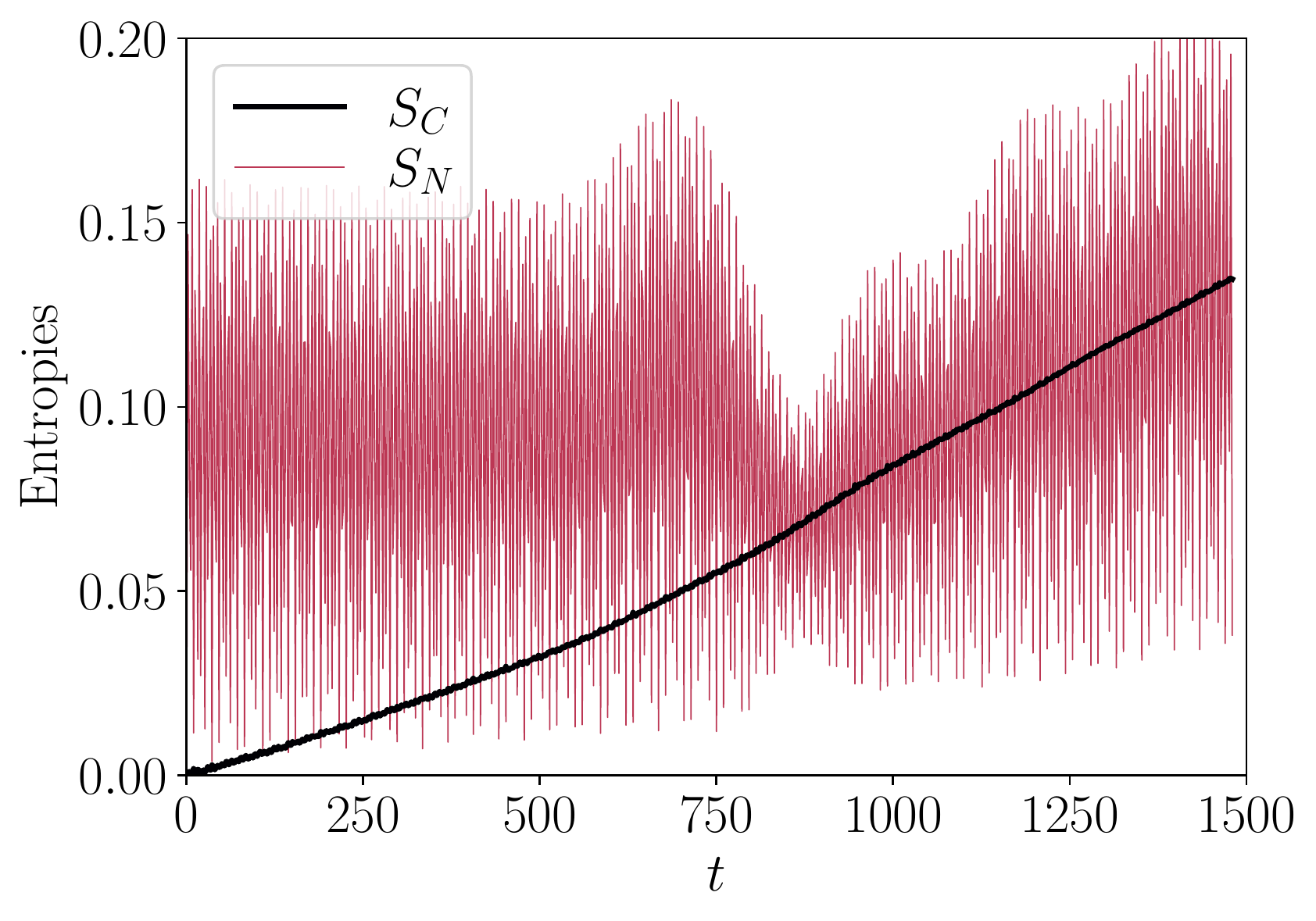} 
\caption{Time dependence of the entanglement entropies in the localized regime. The classical part, the number entropy, reveals fast oscillations corresponding to the single particle transfer across the bond studied.  Configuration (quantum) entropy is growing in time at least linearly with little fluctuations.  Here we choose $L=50$ and $F=6$.
}
\label{entropies}
\end{figure}
For even stronger tilts $F$ the entropy reveals relatively larger fast oscillations as shown in Fig.~\ref{entropies}. 
It is the classical number entropy which is responsible for oscillations in the localized regime.  It is thus related to a transfer of a single particle across the analyzed bond  and is similar in its origin to Bloch oscillations.  On the other hand,  the quantum part of the entanglement entropy, the configuration entropy, shows smooth practically linear increase in time
with a slight bending upwards.  Note that for large $F=6$ value shown, this growth is quite small in the absolute sense.

\subsection{Quantum quench}
\label{quench}

Let us consider a different scenario of initial state preparation. Instead of initial separable state we consider a quantum quench approach \cite{Naldesi16}. We prepare the ground state of the system for some initial parameters and then rapidly change them to the final values at which the time-evolution takes place. By a judicious choice of quenched parameters one may probe different energies of the final Hamiltonian.
The method has been shown to be effective in a variety of models \cite{Naldesi16,Yao20,Kubala21} allowing to pin down the possible existence of mobility edges.

\begin{figure}
 \includegraphics[width=\linewidth]{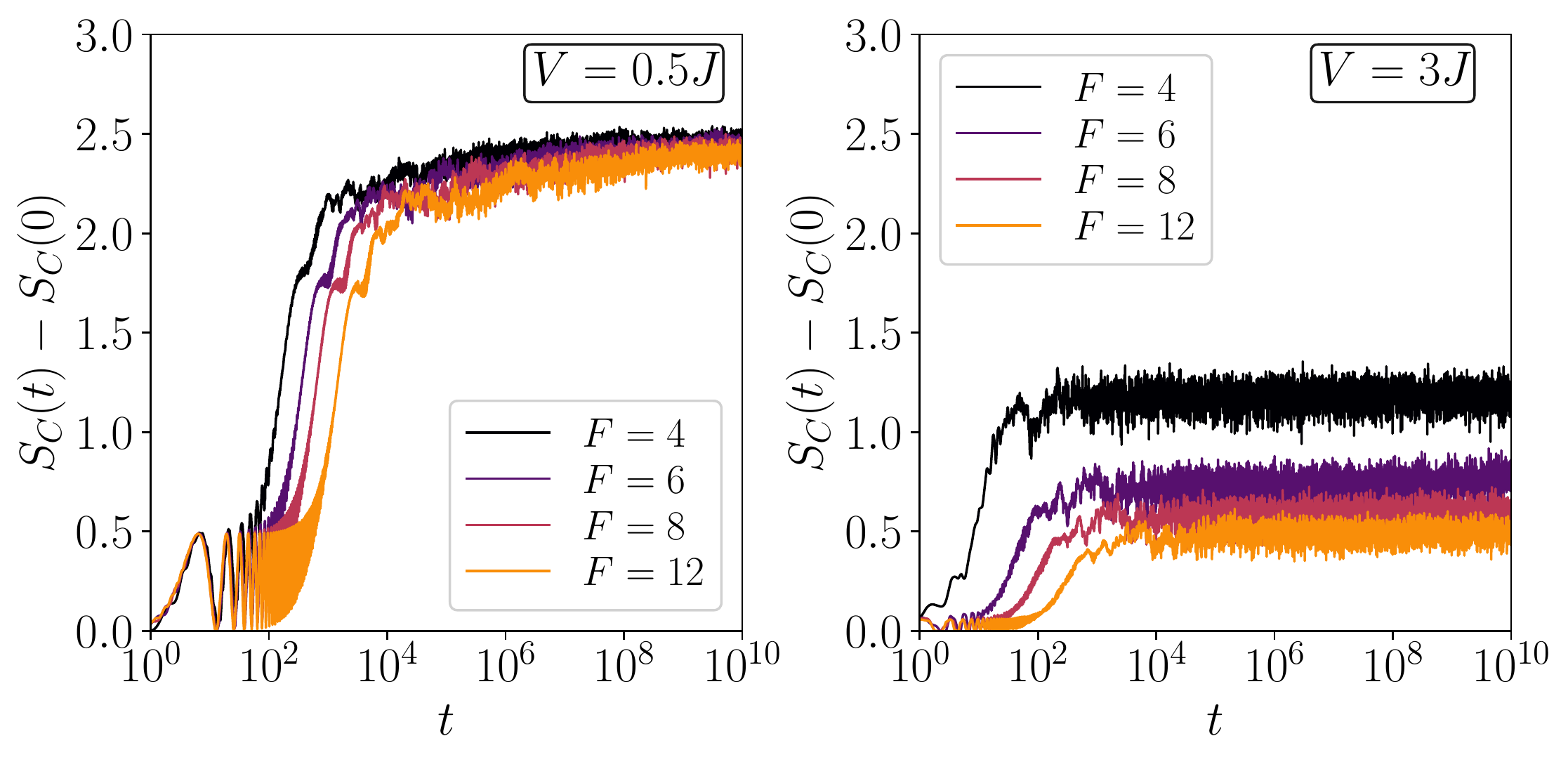}
   \caption{The configurational entropy growth for quantum quench spectroscopy: $V = 0.5J$ (left) and $V = 3J$ (right). The dynamics are entirely different reflecting entanglement properties of the state being quenched - {gapless XY phase} for $V=0.5J$ and antiferromagnetic for $V=3J$. 
    \label{Entropy}
 }
\end{figure}

In the present situation we prepare the system, at some given interaction strength and in the {\it absence} of any tilt in the ground state of
Hamiltonian \eqref{eq: XXZ}. Then we rapidly  turn on the tilt  and observe the time dynamics.
The ground state of XXZ depends on $V$: for $V > J$, the interaction energy dominates and ground state is of a {gapped} antiferromagnetic order while for $V < |J|$ the ground state is in critical XY phase.  The ground state itself has non-vanishing entanglement because of non-zero interaction, and we study the change of configurational entropy after the quench. The growth of $S_C(t)$ brings us the information of how quantum entanglement further builds up in the system. 

Consider the tilt quench (from zero to different final values) for $V=3J$ in which case the initial ground state is antiferromagnetic. While this state is not separable it resembles the N\'eel separable state. The corresponding entanglement entropy growth is thus similar to that observed previously -- compare the right panel of Fig.~\ref{Entropy} where we show the configuration entropy growth from its initial nonzero value. 
The configurational entropy builds slowly initially, and then subsequent rather fast growth ends with a rapid saturation at a fixed $F$ dependent value.
The situation is markedly different for the critical initial state ($V = 0.5J$). Here the initial stage for different $F$ values is similar revealing fast oscillations. As shown in the Appendix \ref{app:qquench} the frequency of these oscillations depends linearly on $V$ indicating that the rearrangement of particles is governed by interactions.  
\begin{figure}
 \includegraphics[width=\linewidth]{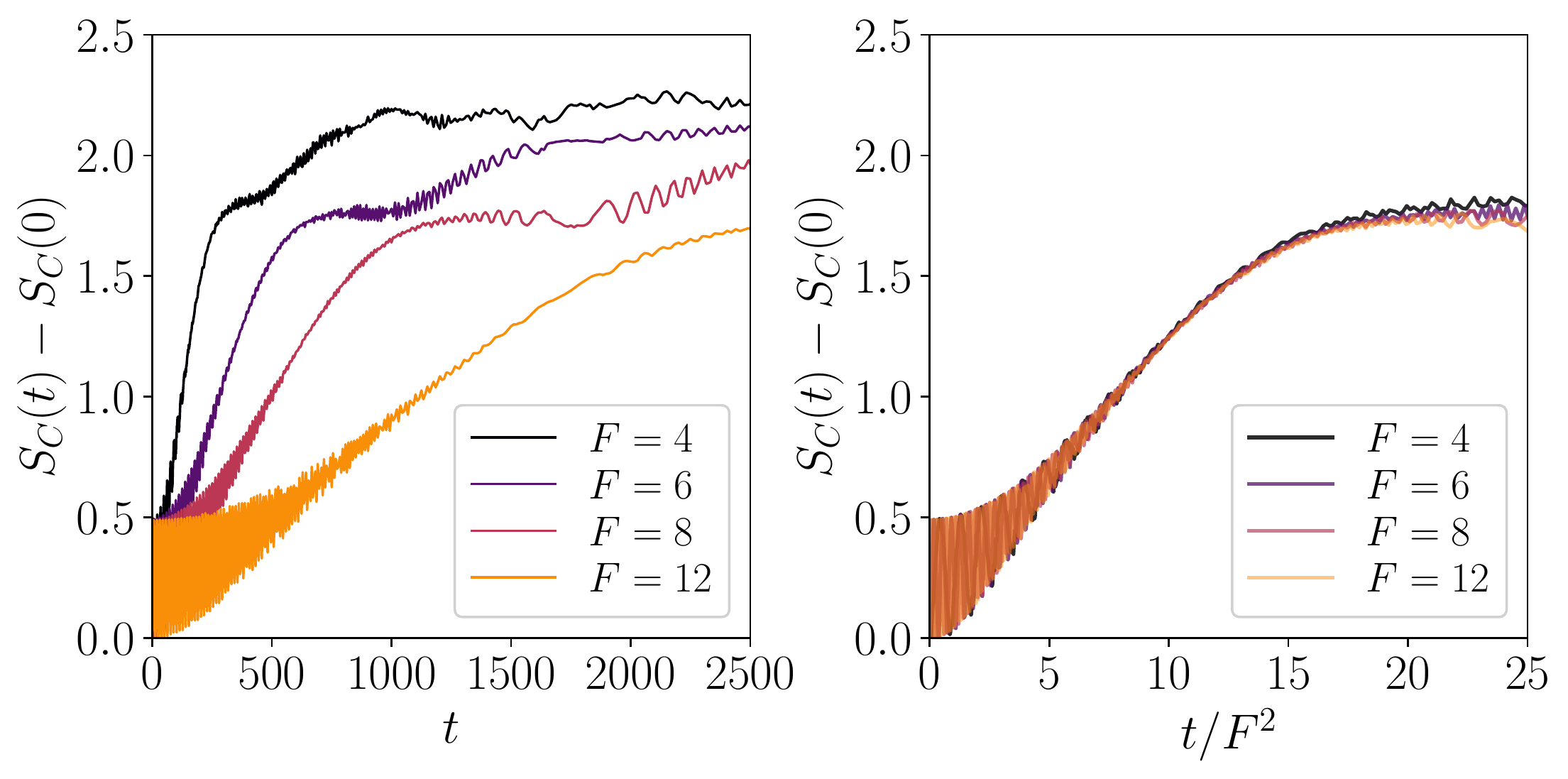}
   \caption{The rapid growing stage of the configurational entropy for $V = 0.5J$. By rescaling the time as $t/F^2$, the configurational entropy falls into a universal curve (right).
    \label{square}
 }
\end{figure}
Beyond the initial oscillation stage, the configuration entropy (we shall consider in detail $V=0.5J$ quench case as entirely different from the N\'eel initial state situation) rapidly grows in $F$ dependent manner. However,  as shown in the right panel of Fig.~\ref{square} this rapid growth falls onto a single curve by rescaling the time by $1/F^2$ factor.  By separately analyzing different $V$ cases (see Appendix \ref{app:qquench} for the evidence) one observes that the correct scaling of time in this region is $t\rightarrow tV/F^2$. 
This numerically identified scaling  is explained extending the mechanism of particle redistributions discussed in detail in \cite{Taylor20}. Consider two particles in a strong field $F \gg V,J$.  For strong tilt the associated energy dominates other energy contributions, the dipole moment of two particles, or its center of mass $x_c = \sum_{i=1,2} x_i/2$ is quasi-conserved. Now take  an initial configuration $|...0110...\rangle$, which appears primarily for smaller interactions $V<1$, being replaced by density wave configurations for larger $V$. The thermalization of states $|...0110...\rangle$ involves spreading out of these two adjacent particles, but under such a strong field $F$, they can not break the dipole conservation. Thus one is left with  the effective tunneling of particle pairs, $\tilde{J}$, that transfers  $|...0110...\rangle \rightarrow |...1001...\rangle$. By second order perturbation theory we obtain
\begin{equation}
\tilde{J} = \frac{J^2}{-F-V} + \frac{J^2}{F-V} \approx \frac{2J^2}{F^2}V
\label{eq:eff}
\end{equation}
providing the explanation for  the $V/F^2$ scaling factor. Higher order contributions, again preserving the dipole moment will lead to
successive orders $F^{-4}$, $F^{-6}$. 
 Since a higher order term has much smaller value, it dominates the dynamics after all dynamics before it is completed. 
 
 \section{Cooperation of the tilted and harmonic potentials}
 
   Consider now the model in which the constant tilt is supplemented by the harmonic potential. As studied in some detail in \cite{Schulz19,Taylor20} the localization in this case is quite similar to MBL. In particular a logarithmic growth of the entanglement entropy is being observed as the additional harmonic term in the potential lifts the partial degeneracy of the spectrum in a pure tilted case. In a different study \cite{Chanda20c} we have considered the dynamics in a harmonic potential alone (with no tilt) finding that 
 localized and delocalized regions may coexist in different parts of the chain, the phenomenon confirmed experimentally recently \cite{Morong21}. Importantly, it turned out that the border between localized and delocalized regions is well predicted involving the notion of the \textit{effective local} field which for the harmonic potential of the form $\mu(i)=A(i-i_0)^2$ equals $F_{\rm loc}=\partial V/\partial i = 2A(i-i_0)$. The transition occurred whenever $F_{\rm loc}\sim1$, separating the regions of smaller and larger local field. 
This behavior is easily understood as many-body localization length in real space is very small, and also for Stark localization it extends over
 few sites only. With the assumed chemical potential in the form $\mu_l=Fl+Al^2/L^2$ the local field becomes $F_{\rm loc}=F+2Al/L^2$
 (note that the curvature of the potential is in our convention equal to $2A/L^2$). 
 
 \begin{figure}
  \label{loc-imb}
 \includegraphics[width=.9\linewidth]{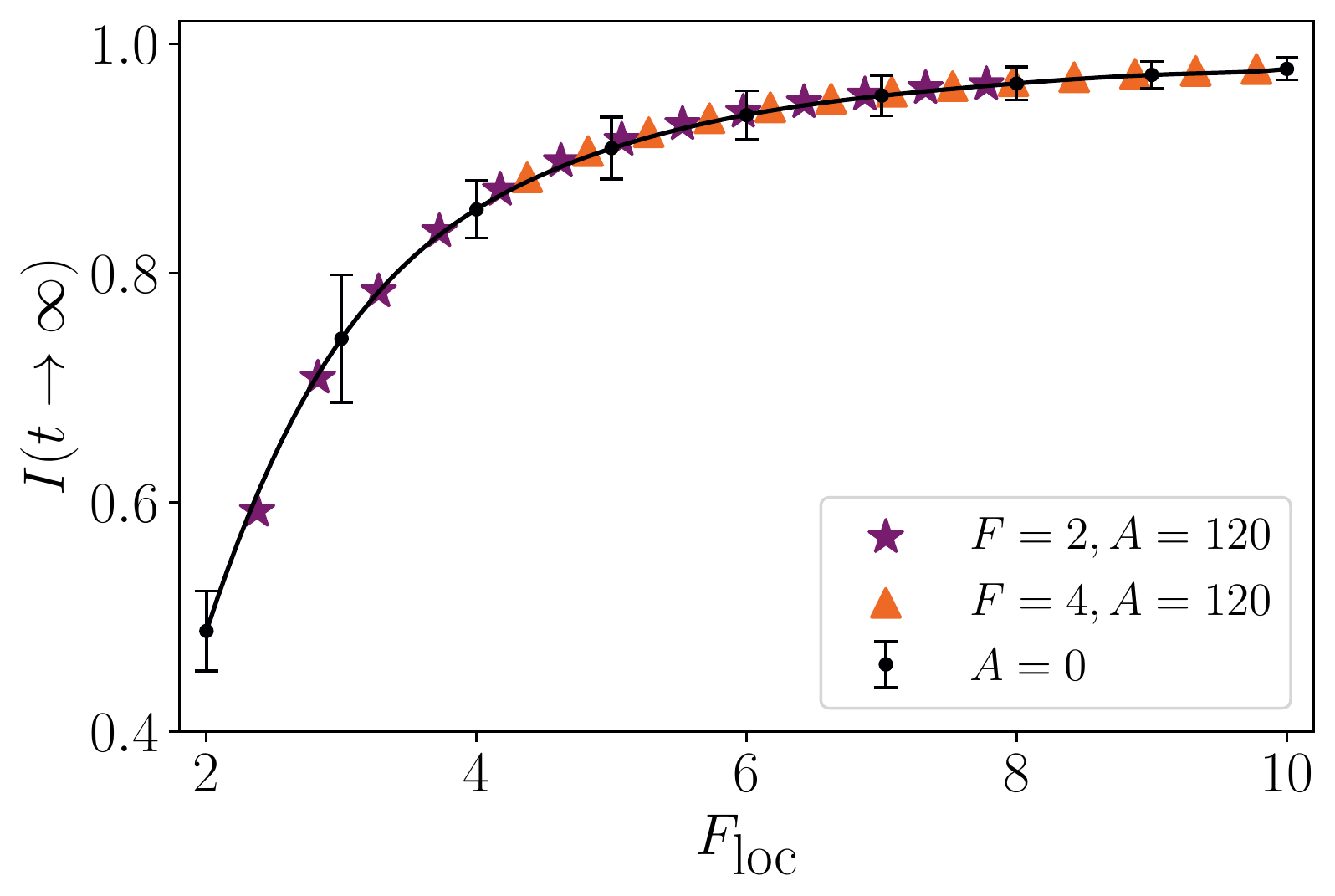}
   \caption{The long-time {\it local} imbalance value (as defined by \eqref{locI}) follows an universal curve dependent on $F_{\text{loc}}$ only. The black circles give reproduce the imbalance from Fig.~\ref{fig:imbalnew1} for $A=0$; the red stars and orange triangles indicate the local imbalance found on bonds corresponding to given $F_{\text{loc}}$. While the global tilt values strongly differ, all points lie on a single curve. 
 }
 \label{fig:imbaloc}
\end{figure} 
For a pure tilt we were able to relate the tilt $F$ with the final imbalance value -- see Fig.~\ref{fig:imbalnew1}. Now, in the presence of the additional harmonic potential the local field changes along the chain and no such a simple connection to the final global imbalance can be made. However, one may define a {\it local} imbalance comparing occupations of just two sites $i$ and $i+1$. For the initial N\'eel state, at one site the spin is up, at the other down (corresponding to occupied and empty sites in the fermionic language) initially, and the change of spin orientation with time may be analyzed. Let us define a local imbalance:
\begin{equation}
I(i)=2 |\langle S^z_i0)\rangle\langle S^z_i(t)\rangle + \langle S^z_{i+1}(0) \rangle\langle S^z_{i+1}(t)\rangle|,
\label{locI}
\end{equation}
centered, to be precise, in the middle of the bond linking site $i$ with $i+1$. With this position we can associate, for a given $F$ and $A$ 
a local field $F_{\text{loc}}$. For $A=0$, $F_{\text{loc}}=F$ and the  local imbalance is the same for all $i$ (again we drop sites very close to the edges of the system) and equals the global imbalance plotted already in Fig.~\ref{fig:imbalnew1}. 

The comparison of local imbalance for different $A$ values is shown in Fig.~\ref{fig:imbaloc} as a function of $F_{\text{loc}}$. Observe that data for different $A$ lie on an universal curve. Thus the local imbalance is solely determined by $F_{\text{loc}}$. This result is even more spectacular keeping in mind the drastically different growth of entanglement entropy for pure tilt and in the presence of an additional harmonic component.

\begin{figure*}
 \includegraphics[width=.65\linewidth]{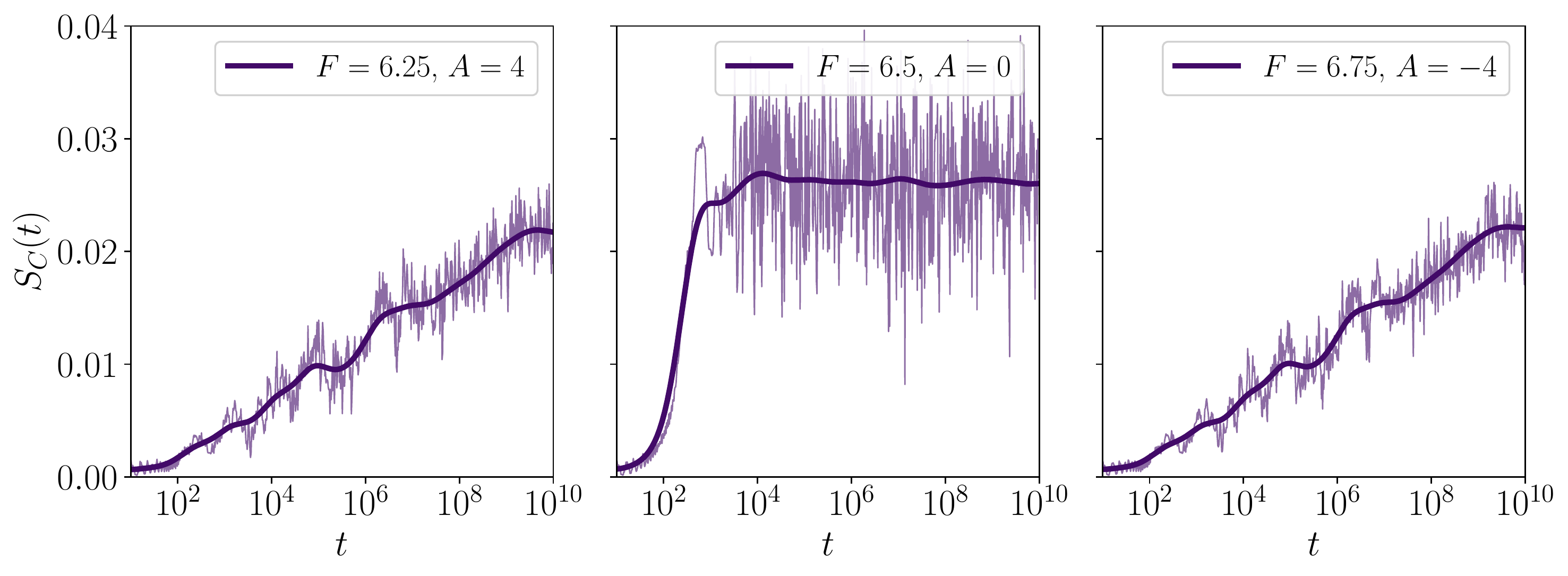}
   \caption{The configuration (quantum) part of the entanglement entropy growth at the central bond for $L=16$. The values of of the tilt $F$ and curvature $A$ are chosen in such a way so the local field $F_{\text{loc}}=F+A/L$ at the middle of the chain remains the same.  Despite that, the entropy growth is different showing that the local
  field notion has a limited applicability. The presence of the harmonic trap clearly changes the fast linear growth followed by saturation into the logarithmic MBL-like growth. Note, however, the similarity of time-dependencies presented in left and right panels - see text for discussion. 
 }
 \label{fig:site}
\end{figure*}
However, the local field approach, while so successful for imbalance must partially fail for the entanglement entropy growth which
substantially differs in the presence or absence of the harmonic component as discussed previously \cite{Schulz19,Taylor20}.
This is further exemplified  in Fig.~\ref{fig:site} for the N\'eel initial state  where we consider configuration (quantum) part of the entanglement entropy, the most relevant one in the localized regime  (recall that the number entropy shows large oscillations - Fig.~\ref{entropies}).
While in the absence of the harmonic potential (central panel) the fast growth is followed by a saturation, the presence of the harmonic
component in the potential makes the growth almost logarithmic in time, with superimposed small oscillations. Note that, surprizingly both left and right panels look very similar despite the fact that the harmonic component has either positive or negative curvature. It seems, that
once we disregard the pure tilt case as special, the local field (same for both panels) provide some information on the configuration entropy growth.

\begin{figure*}
 \includegraphics[width=\linewidth]{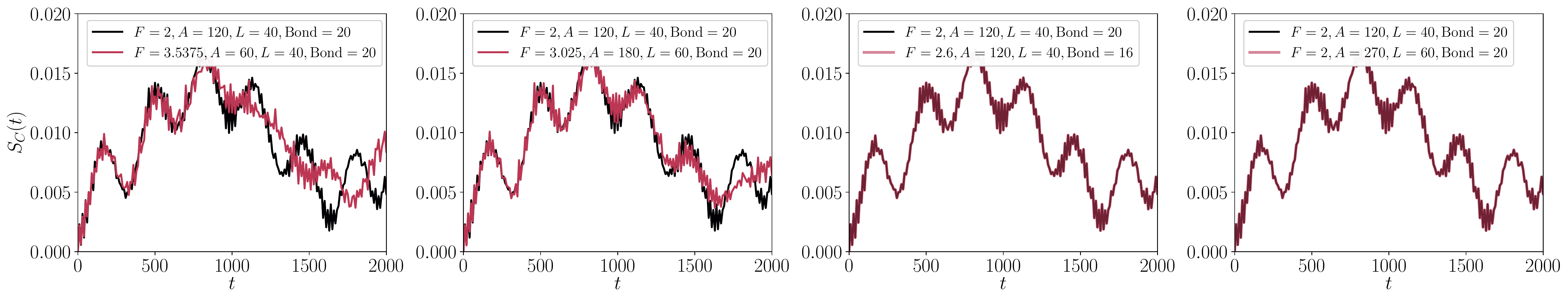}
   \caption{The configuration (quantum) part of the entanglement entropy time dependence for large system sizes. The values of of the tilt $F$ and curvature $A$ are chosen to make the local field $F_{\text{loc}}=F+A(2l+1)/L^2$ the same at the bond, $l$, indicated.  The time-dependence is similar up to $t\approx 1000$.  For longer times the curvature of the potential $A/L^2$ determines the time evolution -- for a discussion see text. 
 }
 \label{fig:site2}
\end{figure*}
This importance of the local field for configuration entropy dynamics is further evidenced in Fig.~\ref{fig:site2}. The configuration entropy 
time dependence for the bond $l=20$ (i.e., the bond between sites $20$ and $21$), $F=2$, $A=120$, and $L=40$ (corresponding to the curvature $\alpha=0.15$ in the lattice spacing units) is compared with similar cases for different tilt and curvature components, as well as different system size but sharing the same value of the local field $F_{\text{loc}}=5.075$ on the analyzed bond. Interestingly, for quite a long time it seems that it is this local field value which determines the entropy dynamics as curves coincide roughly up to $t=1000$ (which seems to be at the edge of experimental feasibility \cite{Scherg20}).
Only for later times the difference appear. They are canceled provided both $F_{\text{loc}}$ and the curvature $2A/L^2$ of the harmonic component are the same. This may be understood in a simple perturbative picture in which the harmonic potential contribution to the dynamics
starts to play some role after a sufficient time. {It is to be noted that in the noninteracting case, i.e., $V=0$, the configurational entropy remains practically  zero for the above choices of parameters.}

 \begin{figure}
 \includegraphics[width=.97\linewidth]{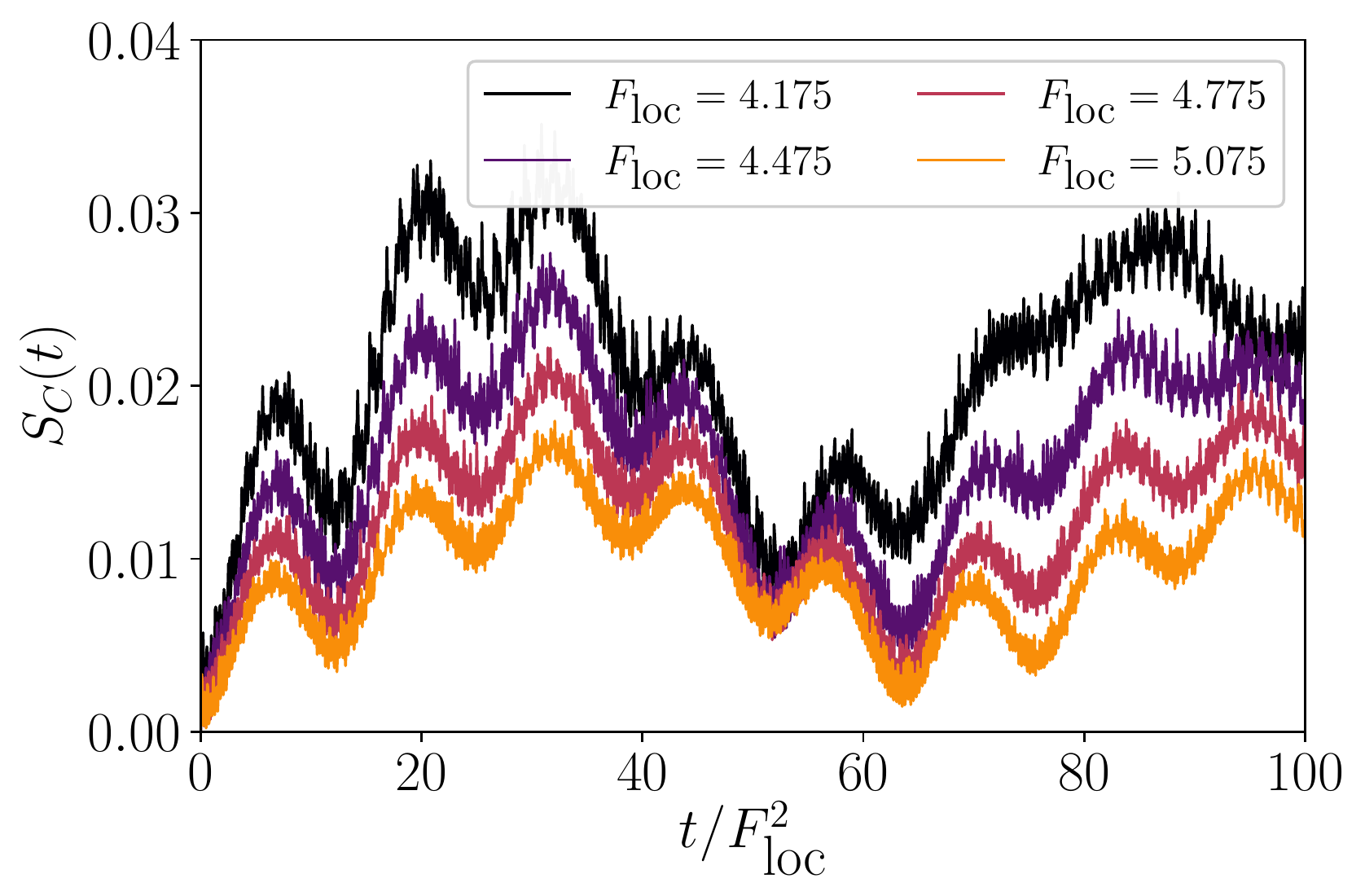}
   \caption{The configuration entropy at 
 different bonds for   $F=2$, $A=120$ and $L=40$ giving different local fields.
    Note that scaling of time with $1/F_{\text{loc}}^2$ unifies the period of oscillations.
 }
 \label{fig:site3}
\end{figure}
The scaling of initial oscillations of the configuration entropy may be identified by looking at different $F_{\text{loc}}$ values. Those can be obtained from a single run with nonzero curvature, we consider the case represented by black curve in Fig.~\ref{fig:site2}. As it turns out the correct scaling of time occurs for $1/F^2_{\text{loc}}$ factor -- see Fig.~\ref{fig:site3}. Let us recall the explanation of the scaling scenario in Stark regime in terms of simultaneous
tunneling of two particles as ``conserving the global dipole moment''. Here apparently we have a similar scenario but due to the presence of the harmonic potential,   $F_{\text{loc}}$ takes the role of $F$.
 
 \begin{figure}
 \includegraphics[width=.97\linewidth]{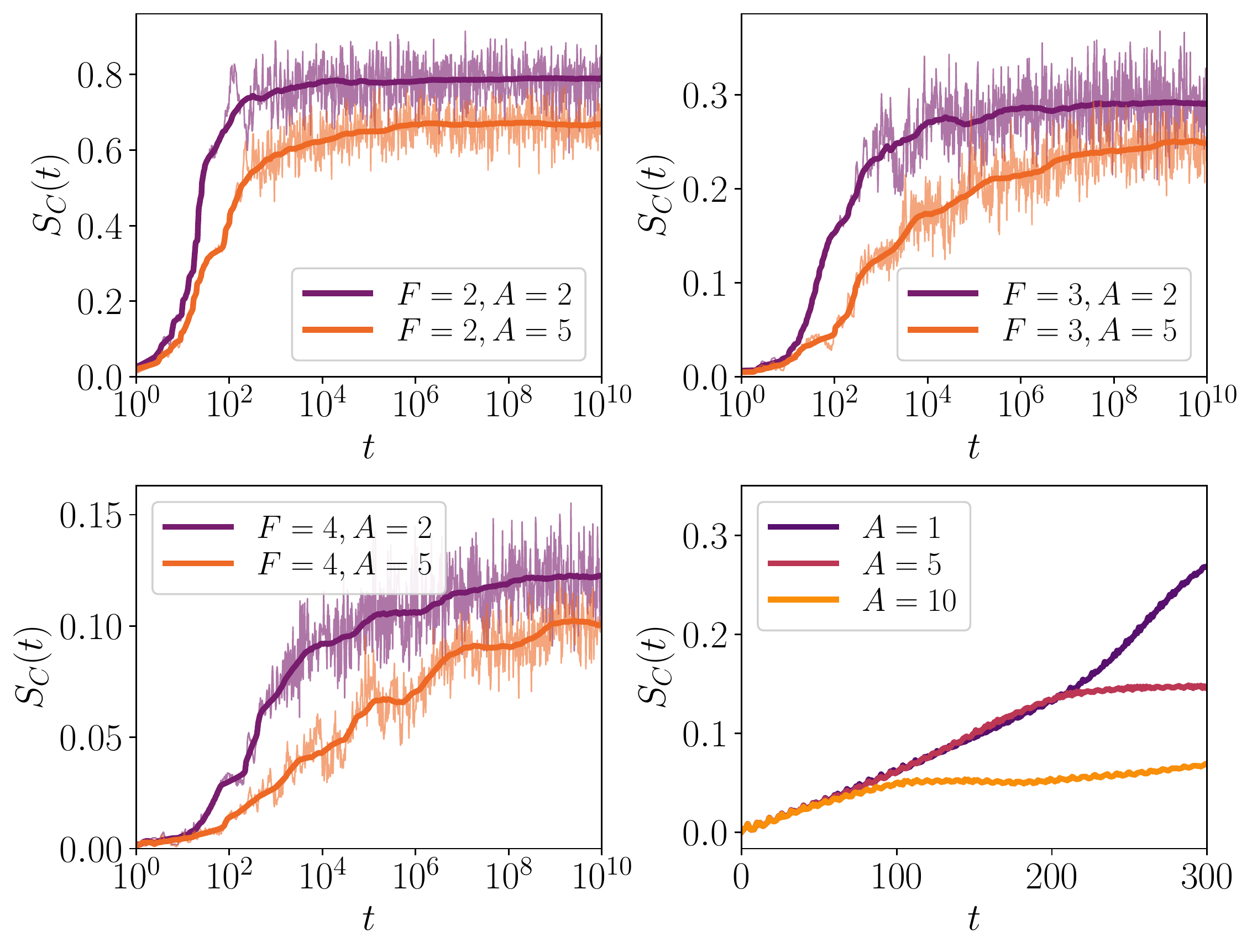}
   \caption{The growth of the configuration entanglement entropy in time for different systems and parameters. Top row as well as bottom left panels correspond to small system size of $L=16$. The entropy growth shows a logarithmic character for a sufficient, large tilt only (bottom left panel). For smaller $F$ (but still in localized regime) one observes a fast growth followed by saturation. The initial evolution, as shown in the bottom right panel for $L=40$, follows the linear growth until the system resolves the energy level splitting due to the presence of a weak harmonic potential. For that it is the curvature of the potential that is relevant - the three different curves correspond to the same $F_{\rm loc}=4$.
 }
 \label{fig:newent}
\end{figure}
As shown in Fig.~\ref{fig:site} in the presence of a harmonic confinement and a large tilt, the long time entanglement entropy growth seems to follow $\ln(t)$ mean growth with additional oscillations which we just analyzed. Similar logarithmic growth has been stressed in earlier studies \cite{Schulz19,Taylor20}. This is not always the case as exemplified in Fig.~\ref{fig:newent}. For a N\'eel initial state  the logarithmic growth
of the configuration (and thus total) entanglement entropy starts to appear for sufficiently large $F$ and $A$. For weaker tilt, as in the top row, small system studies (that enable long time dynamics studies) indicate that the initial, rather fast entropy growth is followed by a saturation as for pure tilt case. Apparently, for $F=2$ lifting of degeneracies induced by a harmonic field is not sufficient. During the short initial time the system does not resolve individual levels and the entropy grows in a way resembling a pure linear tilt. Then the saturation of the entanglement entropy growth due to $F$-field localization prevents further $S_C$ growth. With increasing $F$ the transition to a more ``standard'' logarithmic growth is observed for sufficiently large $F$ and $A$ (bottom left panel of Fig.~\ref{fig:newent}).

Our findings in terms of entanglement entropy growth are apparently at variance with results reported in \cite{Schulz19,Taylor20} that stress the similarity of SMBL in additional harmonic potential and the standard MBL, in particular showing the logarithmic entropy growth. This discrepancy may be due to different initial states. We considered experimentally relevant N\'eel initial state while, e.g., \cite{Schulz19}
averages over different initial states empty close to the bond on which entropy is calculated. Averaging over different initial states with
significantly different dynamics may result in the average entropy growth different from that observed for the N\'eel state. This aspect requires a separate detailed study.

The bottom-right panel illustrates the time dynamics for short times (but for larger $L=40$ system) exemplifying the initial linear growth associated to degeneracies of the spectra for pure linear tilt \cite{Schulz19,Taylor20}. With increasing curvature the linear growth is followed for a shorter time, while the system realizes that the degeneracy is lifted, the growth becomes slowed and undergoes oscillations discussed above.

\begin{figure}
\includegraphics[width=\linewidth]{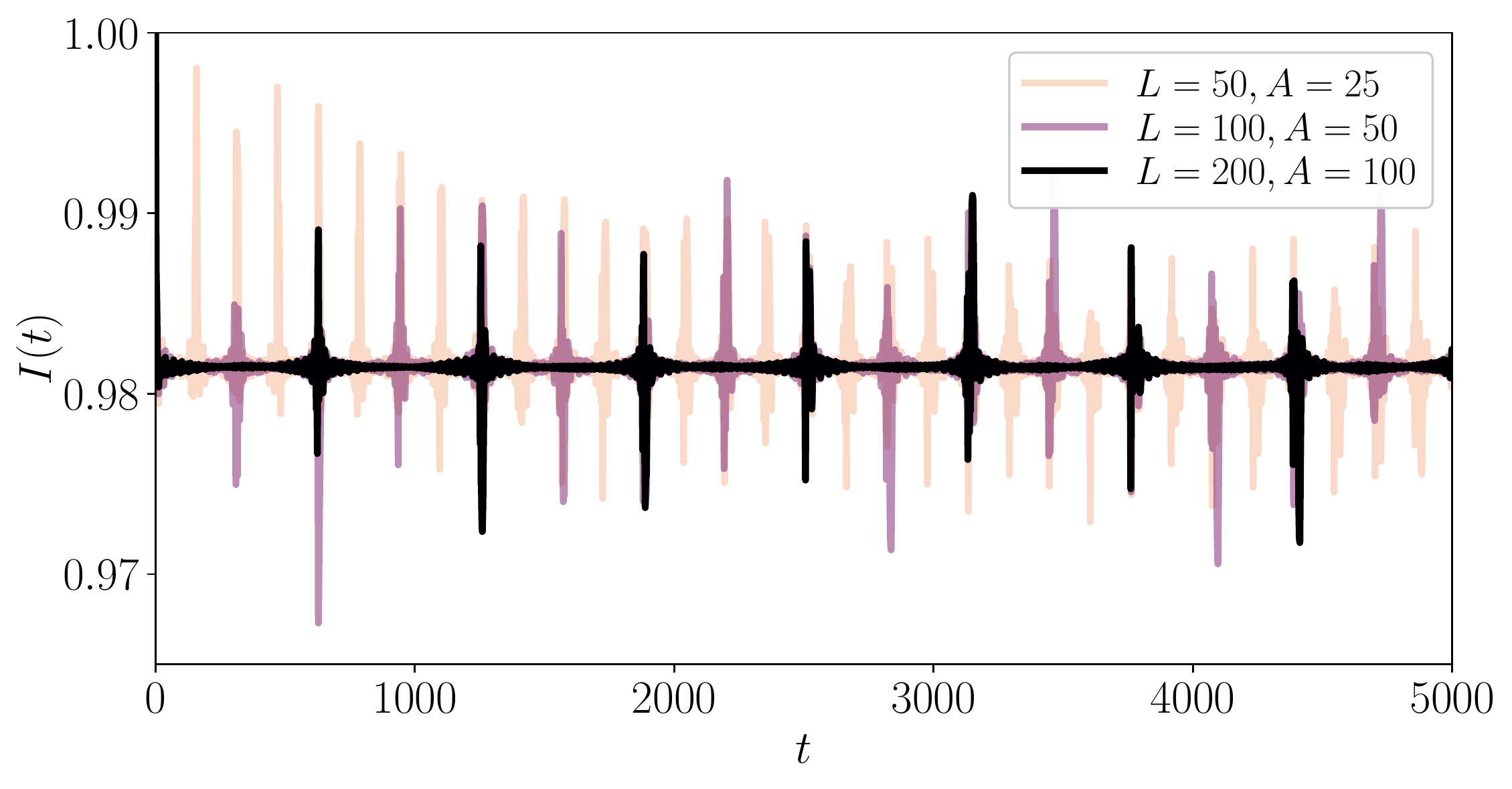}
\caption{The imbalance  for large $F=10$ and the additional harmonic potential  saturates at a constant value revealing,  however,  pronounced revivals that occur at the period $T=\pi L^2/2A$ as shown for different system sizes.  The revivals are due to splitting of the (quasi-)degenerate total dipole manifold in the presence of a harmonic addition to the potential.  For further discussion see text.}  
\label{fig:ImbH}
\end{figure}
Lastly, let us consider in detail
 the time dependence of the imbalance for different system sizes and different curvatures of the harmonic potential for even larger tilts.  For $F=10$ the imbalance saturates very fast at the high value about 0.98 showing,  however,  regular revivals.  We have verified that the period of the revivals does not depend of $F$,  the resulting curves are similar for $F=6$ or $F=4$  at different  saturation level.  For each system size the $A$ value is chosen in such a way that a local field at the end 
of the chain is increased from $F_{\rm loc}=10$ to $F_{\rm loc}=11$. 
To a good precision  the period of revivals
equals $T=\pi L^2/2A$ as if corresponding to equally spaced levels with spacing $\omega=2\pi/T=4A/L^2$. 
Indeed by a direct evaluation (calculations were performed at small $L=12, 14$ sizes) one may check that the (quasi-)degenerate (for $A=0$) manifold of states having the same global dipole moment ${\cal D}$ splits in the presence of the additional harmonic potential of curvature $A$ into the set of almost equally spaced levels with the spacing $\omega/2$.  However, only every second state in this manifold has a non-zero overlap with the initial N\'eel state contributing to the dynamics -- thus indeed the relevant spacing is equal to $\omega$. 
Another important point to note is that since the emerging dipole conservation becomes exact only in the large $F$ limit, 
such revivals with period $T=\pi L^2/2A$ become more pronounced with increasing $F$.


\section{Conclusions}

We performed a detailed analysis of localization in disorder-free potentials considering both the pure static field and the system with an additional weak harmonic potential. We have studied both small system sizes where exact diagonalization and Chebyshev propagation enables accurate results for very long times, as well as large systems have been studied with the help of matrix product states algorithms. 
By a detailed analysis of the entanglement entropy we have verified that the character of localization in a pure static field is different from standard disorder induced MBL, in particular a fast initial growth of the entanglement entropy in a static field followed by a full saturation is different from the logarithmic growth known from MBL. Similarly while the number entropy in the standard MBL seems to grow as $\ln(\ln(t))$ \cite{Kiefer20,Kiefer21} at long times - we observe strong fast Bloch-like oscillations, but filtering them out we observe no trace of the slow  $\ln(\ln(t))$-like growth.
The rapid initial growth of the entanglement entropy can be associated with a rapid coherent coupling within the degenerate manifold for the pure tilt \cite{Schulz19,Taylor20}. The large system size data provide a further confirmation for these claims. 

Apart from the time dynamics of the initial N\'eel state we considered also a quantum quench scenario. There we have identified the mechanism of the entanglement entropy growth in the localized regime as an effective second order tunneling of two particles conserving the emergent global dipole moment.

In the presence of an additional harmonic potential the character of the dynamics changes. While the resemblance to the standard 
 MBL was noted  in \cite{Schulz19,Taylor20} in, e.g., the logarithmic growth of the entanglement entropy  in time, our simulations for the N\'eel initial state show this to be the case only for sufficiently large tilt and curvature of the harmonic potential. This observation is limited to small systems only as it requires long evolution times. For shorter, experimentally reachable times we analyze the large period oscillations of the configuration entropy and observe there the characteristic time scales related to the {\it local} tilt values (at the bond where entanglement is analyzed).  The detailed analysis of the imbalance from the initial N\'eel state as well as the entanglement entropy dynamics indicates that while ``local'' field approximation introduced by us in \cite{Chanda20c} may provide very good estimate for the system behavior,  it fails for long times in providing a consistent interpretation of the entanglement entropy time dynamics. There an additional influence of the potential curvature must be taken into account to explain the simulation results.  Studying large systems we have shown that breaking of the approximate global dipole conservation may also result, in the strongly localized regime, in the appearance of the characteristic revivals in the long time imbalance.
 
 Finally we would like to stress that the word of caution is necessary. While for small systems we may access large times by exact propagation we cannot do so for larger systems where we use TDVP. So the physics of large systems and long times in tilted potentials
may hide additional surprizes. In particular, in the limit of large tilt we may expect to observe unusual thermalization due to fractonic dynamics
 \cite{Nandkishore18,Pai19,Pai20,Pretko20,Taylor20,Khemani20} which is, however, beyond the scope of this work.

 \begin{acknowledgments}
This work benefited from the invaluable input of Dominique Delande who contributed dominantly to the TEBD parallelized code used in this work.  
Equally important were discussions with Piotr Sierant on various aspects of the physics of tilted models. We acknowledge both contributions.
J.Z. is grateful to Pablo Sala for suggesting the comparison of imbalance to the noninteracting case. Discussion with Frank Pollmann was also very enlightening.
The numerical computations have been possible thanks to  High-Performance Computing Platform of Peking University as well as PL-Grid Infrastructure.
The TDVP simulations have been performed using ITensor library (\url{https://itensor.org}).
This research has been supported by 
 National Science Centre (Poland) under projects  2017/25/Z/ST2/03029 (T.C.) and  2019/35/B/ST2/00034 (J.Z.)
 \end{acknowledgments}

\appendix

%

\section{Remarks on matrix product states based time dynamics for large systems}
\label{app:mps}

The study of larger systems is possible entirely within the necessarily approximate time-dynamics schemes such as time-evolving-block-decimation (TEBD) \cite{Vidal03},
time-dependent density matrix renormalization group tDMRG \cite{White04}, or more recent
numerical schemes based on time dependent variational principle (TDVP)  \cite{Haegeman11, Haegeman16, Paeckel19}. 
The early studies involved TEBD \cite{Vidal03,Daley04,Gobert05,Kollath05} that allowed for relatively efficient simulation of interacting systems also in the context of MBL   
\cite{Hauschild16,Sierant17,Sierant18,Zakrzewski18}. 
Recently
TDVP was used by the Karlsruhe group to estimate the onset of MBL in large systems \cite{Doggen18,Doggen19} -- for a discussion see \cite{Chanda20t}.  The latter work critically inspected applicability of TEBD and TDVP for accurate description of time-dynamics in systems of realistic size. 
In this work, we mostly use a \textit{hybrid} variation of the TDVP scheme  \cite{Paeckel19, Goto19,Chanda20, Chanda20t}, where we first use two-site version of TDVP.  At this stage  the (auxiliary) bond dimension grows up to a prescribed $\chi_{max}$ value.  Reaching $\chi_{max}$,  we switch to the one-site version (avoiding  errors due to a truncation of singular values  in the two-site version \cite{Paeckel19, Goto19}). 
Typically for large $F$,  in the localized regime, we often do not reach $\chi_{max}$ so the simulations are fully converged (up to numerical errors).  For TEBD we use an efficiently parallelized home made
package that utilizes the recent breakthrough for Lie-Trotter-Suzuki decomposition \cite{Barthel20}. .
Disappointingly,  the execution times needed for propagation using TDVP of realistic system sizes in a static tilt case were found empirically to be quite long compared to disordered scenarios.  
This may be related to the fact that in presence of static fields the energy levels of the Hamiltonian are spread covering a broad range of energies and local Lanczos exponentiations in the TDVP scheme require large Krylov space to converge. 
On the other hand the simulations, in the absence of disorder, require just a single run, with no disorder averaging. That, with a necessary patience, allows us to
study system sizes and times unprecedented until now. 

\begin{figure}
 \includegraphics[width=.9\linewidth]{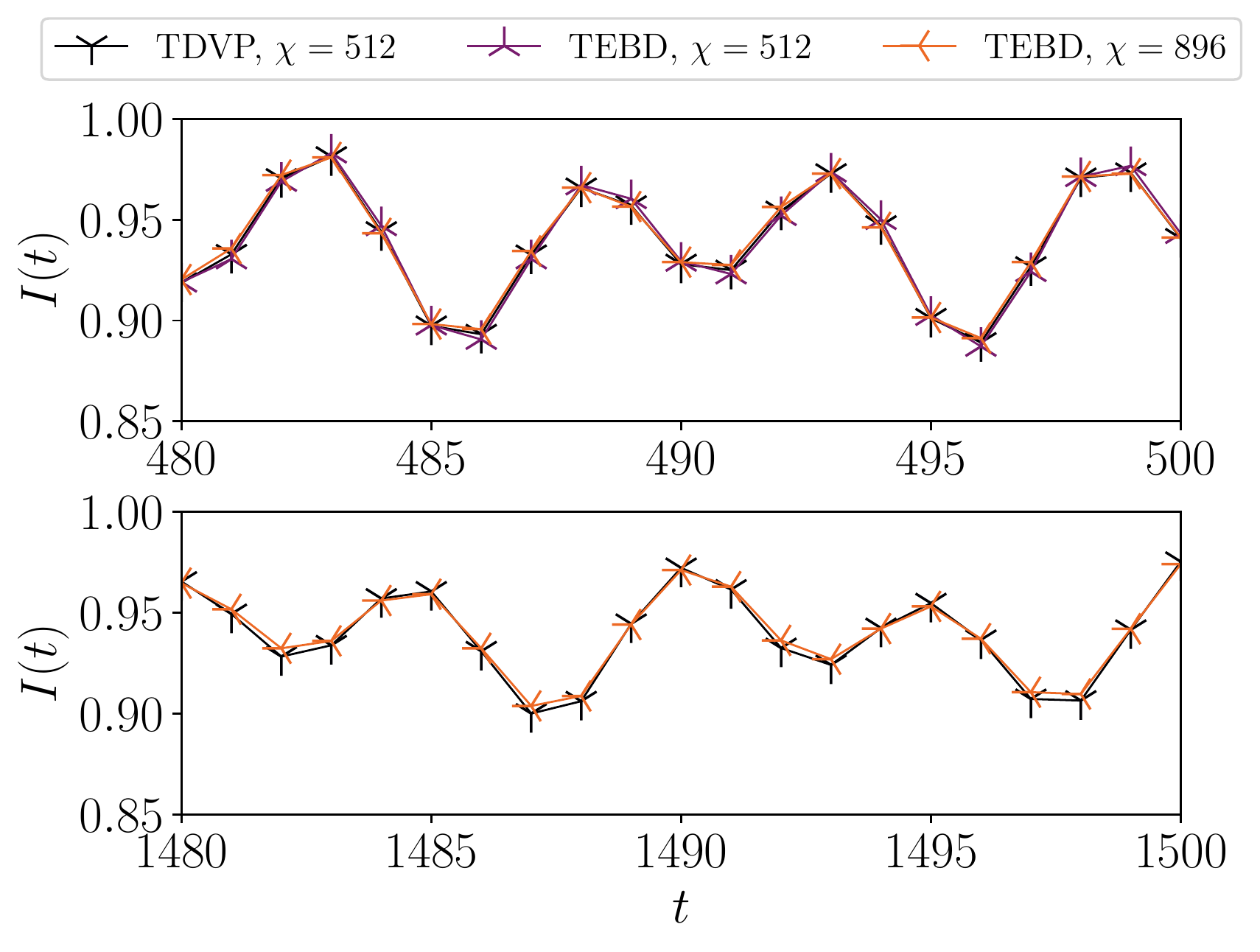}
   \caption{The time dependence of the imbalance for $F=6$, i.e., in the localized regime, obtained with both TEBD and TDVP algorithms for the spin chain of length $L=40$. For comparison short time stretches are shown only, the numerically obtained values are connected by straight lines.  The agreement is spectacular indicating
   the convergence for times reaching the experimental limits of preserving coherence in systems studied. The auxiliary space dimension $\chi_{max}=512$ for TDVP and $896$ for TEBD system - the latter requires larger auxiliary space for the similar convergence at long times. For intermediate times  $\chi_{max}=512$ is sufficient for TEBD too (compare the top panel).
 }
 \label{fig:tebd}
\end{figure}
To test the accuracy of our studies we compare the simulations performed with TDVP, where due to the sweeping procedure, an efficient parallelization of the code seems difficult, with TEBD where such a parallelized code with higher order Suzuki-Trotter scheme is possible. Fig.~\ref{fig:tebd} shows that the imbalance evolution in time for a state in the localized regime is essentially the same using TDVP and TEBD propagation schemes. Since, as shown in \cite{Chanda20t} TEBD {\it overestimates} the imbalance while TDVP {\it underestimates} it, the agreement of both schemes assures that the results obtained are fully converged. Note that even deep in the localized regime we have to use slightly larger auxiliary space dimension for TEBD, as described in the caption of Fig.~\ref{fig:tebd} - the possible growth in the execution time is compensated by efficient parallelization possible for TEBD.
Note also that we do not observe any sign of the accumulated Trotter errors which haunts TEBD implementation \cite{Daley04} -- presumably because of high order decomposition used \cite{Barthel20}.

\section{The frequency filter}
\label{app:filter}

To minimize the effects of high frequency oscillations in the time signal of a particular observable, say $O(t)$, we first perform the fast Fourier transform (FFT) in the time domain as $O(t) \xrightarrow[]{\text{FFT}} O(\omega)$. Then we perform a Gaussian convolution to the Fourier transformed data $O(\omega)$ as $O(\omega) \rightarrow \tilde{O}(\omega) = O(\omega) \exp\left(-\frac{\omega^2}{2\sigma^2}\right)$ for reducing the contributions coming from higher frequencies. Finally, an inverse fast Fourier transform (iFFT) is performed to get the frequency filtered data in the time domain as $\tilde{O}(t) \xleftarrow{\text{iFFT}} \tilde{O}(\omega)$. In our calculations, we have used NumPy's standard FFT and iFFT functions (\url{https://numpy.org/doc/stable/reference/routines.fft.html}), and have chosen the standard deviation $\sigma$ in the Gaussian convolution between $[0.05, 0.1]$ depending on the scenarios.

\section{Details on quantum quench properties}
\label{app:qquench}

\begin{figure}
 \includegraphics[width=0.8\linewidth]{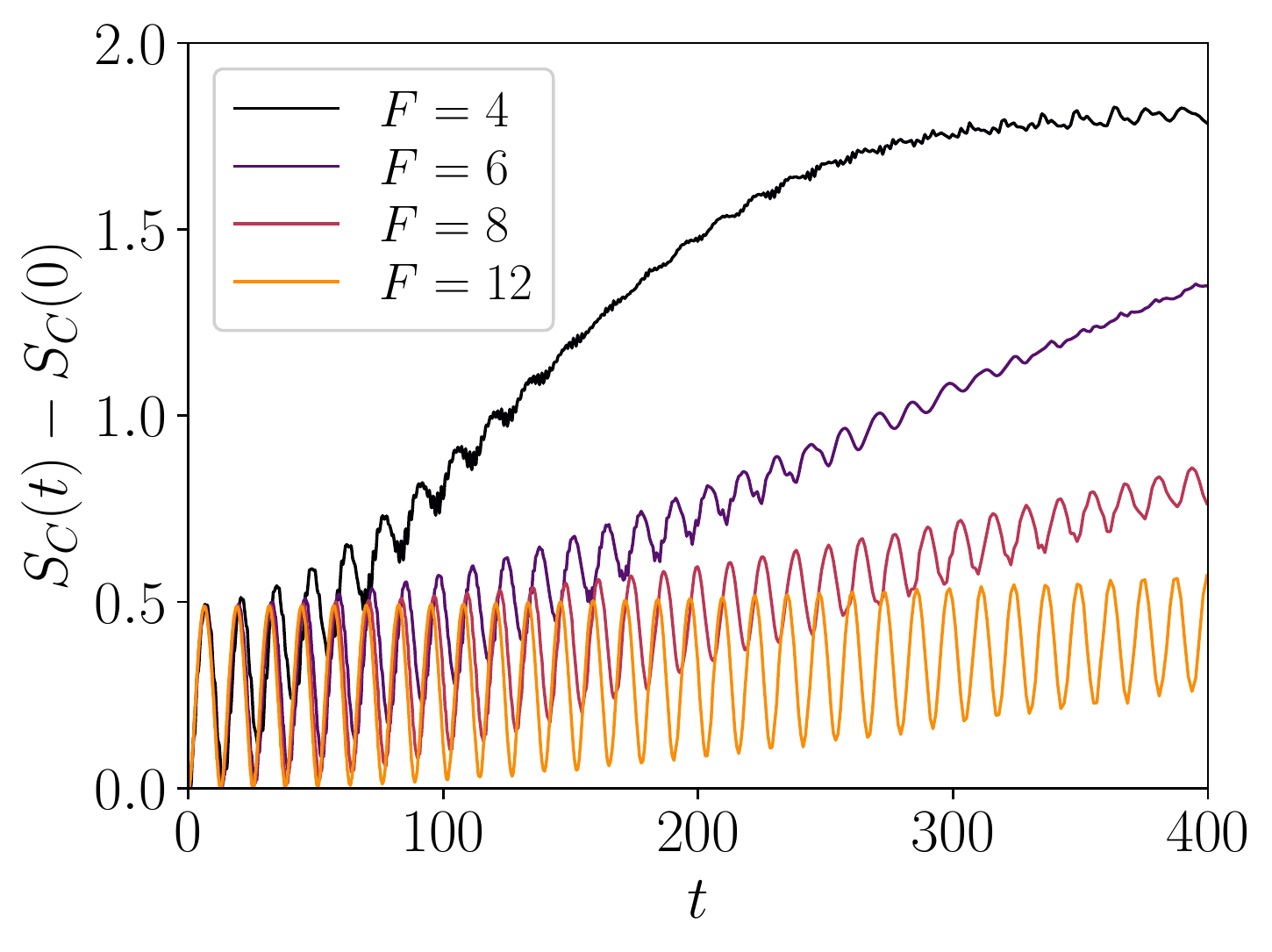}
   \caption{The initial stage of configurational entropy growth of $V = 0.5J$. An oscillation with similar frequency with a succeeding linear growing trend.
    \label{linear}
 }
\end{figure}

\begin{figure}
 \includegraphics[width=\linewidth]{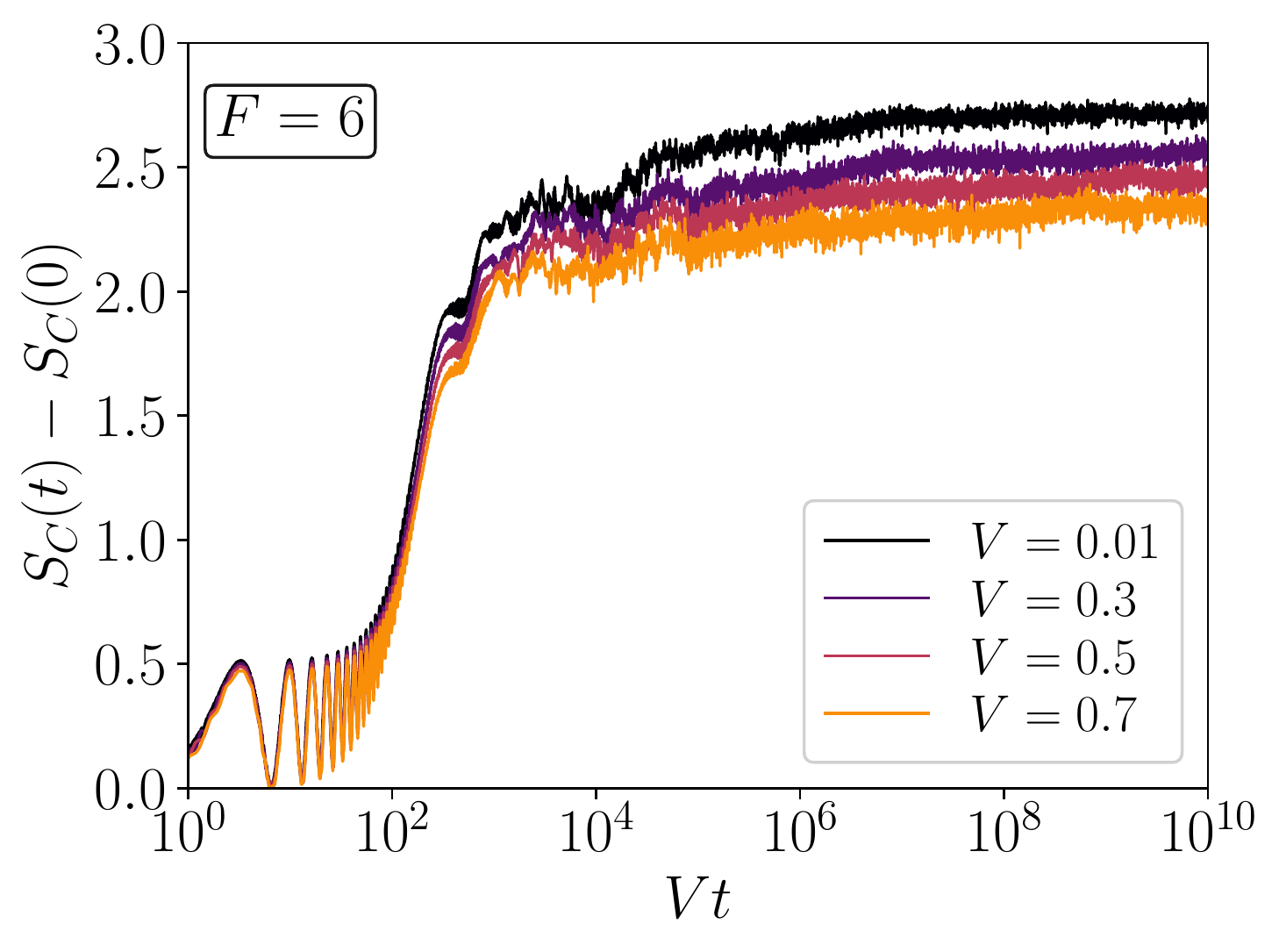}
   \caption{Rescale of $t \rightarrow Vt$ results into the alignment of different growing stages. Thus the effective tunneling should be linear in $V$.
    \label{inter}
 }
\end{figure}

We take a closer look at weak interaction case by examining different time scales. The initial stage is an oscillation accompanied by succeeding linear growth, see Fig.~\ref{linear}. After this stage, there is a rapid growth before the much slower growth occurring for long times. The initial oscillation is $F$ independent, we see in Fig.~\ref{inter} that rescaling the time $t \rightarrow Vt$ for different $V$ values at a given fixed  $F$ makes the oscillations and the subsequent rapid growth coalesce. Thus, combining with the evidence presented in Fig.~\ref{square}, the dynamics is dependent on the single combined parameter $V/F^2$ in agreement with the effective pair tunneling explanation Eq.~\ref{eq:eff}.

%


\end{document}